\documentclass{aa}
\usepackage{times}
\usepackage{graphics}
\usepackage{h1539}

\begin{document}
\setlength{\arraycolsep}{0pt}
\thesaurus{06(02.08.1; 08.02.1; 08.15.1; 08.18.1)} 
\title{Tidal evolution of eccentric orbits in massive binary systems}
\subtitle{a study of resonance locking}
\titlerunning{Evolution of eccentric orbits}
\author{M.G.~Witte \and G.J.~Savonije}
\offprints{M.G.~Witte}
\mail{marnix@astro.uva.nl}

\institute{Astronomical Institute `Anton Pannekoek', University of
  Amsterdam, Kruislaan 403,  NL-1098 SJ Amsterdam, The Netherlands}

\date{Received 10~May~1999 / Accepted 6~July~1999}
\maketitle
\begin{abstract}
  We study the tidal evolution of a binary system consisting of a $1.4\,\mathrm{M}_\odot$
  compact object in elliptic orbit about a $10\,\mathrm{M}_\odot$ uniformly rotating main
  sequence (MS) star for various values of the initial orbital parameters.  We apply our
  previously published results of 2D non-adiabatic calculations of the non-radial g- and
  r-mode oscillations of the uniformly rotating MS star, and include the effects of
  resonant excitation of these modes in the tidal evolution calculations.  A high orbital
  eccentricity enhances the effectiveness of the tidal interaction because of the large
  number of harmonic components of the tidal potential and the reduced orbital separation
  near periastron.  By including the evolution of the MS star, especially of its rotation
  rate, many resonance crossings occur with enhanced tidal interaction.  We analyse the
  phenomenon of resonance locking whereby a particular tidal harmonic is kept resonant
  with a stellar oscillation mode by the combined action of stellar evolution and other
  tidal harmonics.  Resonance locking of prograde g-modes appears an effective mechanism
  for orbital circularization of eccentric orbits.  We consider the orbital evolution of
  the binary pulsar \object{PSR J0045-7319} and conclude that resonance locking could
  explain the observed short orbital decay time of this system if the B-star spins in the 
  direction counter to the orbital motion. 
\end{abstract}


\section{Introduction}
\label{sec:intro}

Two (or more) celestial bodies which orbit each other will experience periodic
deformations due to the fluctuations of the mutually attracting gravitational fields.
The tidally induced flows in these bodies are subject to dissipative processes, causing
a phase lag to develop between the perturbing potential and the displaced material.
The resulting torque changes the orbital parameters of the system, and usually acts to
drive the system towards circularization of the orbit and corotation of the components.
If a binary system is in such a tidally relaxed state, it is relatively simple to
calculate the tidally induced static shape of the components, the so-called equilibrium
tide.  However, in general the tidal deformation is time dependent and requires
dynamical effects to be taken into account.  Cowling \cite*{C41} noted that the tidal
forces can excite gravity (g-) modes in the radiative layers of binary stars, the
so-called dynamical tides.  In case of a high mass early type star the stellar envelope
is radiative and gravity modes can be excited for which significant damping occurs in a
layer beneath the stellar surface, where the local thermal timescale is comparable to
the period of oscillation.  Zahn \cite*{Z77} developed an asymptotic theory valid for
low forcing frequencies where the radiative damping is severe and resonances with
g-modes are smeared out.  However, when the system is far from equilibrium, tidal
forcing contains relatively high frequencies, so that resonances with relatively weakly
damped low radial order g-modes are likely to occur.  Zahn's theory may predict tidal
timescales orders of magnitude too small when the asymptotic low frequency conditions 
are not met.

In a series of papers Savonije \& Papaloizou \cite*{SP83,SP84} presented non-adiabatic
linearised calculations for perturbed spherical stellar models, enabling the study of
resonantly excited g-modes at different stages of evolution of the star.  To include
effects of stellar rotation on the oscillations, an implicit 2D hydrodynamics code has
been developed \cite{SP97} in which centrifugal distortion is discarded, but which
fully accommodates the arising Coriolis forces in the rotating star.  Due to the
Coriolis force the response of a rotating star to a spherical harmonic perturbing force
is itself not a simple spherical harmonic function, but may be numerically solved on a
2D $(r,\vartheta)$ grid.  Using the 2D-code, the effects of rotation on the stellar
(g-)modes can be examined, while also it is possible to study the stellar
quasi-toroidal r-modes \cite[see]{PP78} in detail.  In a previous paper \cite{WS99}
(Paper~I) we utilised the above mentioned 2D code to calculate the g- and r-mode
spectra of a $10\,\mathrm{M}_\odot$ stellar model with a central hydrogen abundance of
forty percent rotating uniformly at speeds up to forty percent of breakup.  In order
to study effects of the star's evolution, we extend the data of Paper~I with some
calculations for a $X_\mathrm{c}=0.2$ stellar model, the results of which are listed in
the appendix of this paper.  The strength of the r-mode resonances, as given by the
area of the peaks, becomes larger in a more rapidly rotating star.  For rapid stellar
rotation with $\Omega_s=0.4\,\Omega_\mathrm{c}$, the strongest r-modes have peak areas
comparable to g-modes with approximately 7--10 radial nodes.  Additionally the long
wavelength weakly damped r-mode resonances occur at quite low forcing frequencies in a
region where the stellar g-modes are severely damped as a result of their short
wavelength.  If, therefore, a star is forced with strong harmonics that lie in this
region, resonant excitation of r-modes can be expected to dominate the orbital
evolution of the system.

The recent discovery \cite[see]{KJBM94} of the binary radio pulsar \object{PSR
  J0045-7319}, which contains a B-star and a neutron star in a highly eccentric orbit,
and for which radio pulsar timing has lead to a fairly accurate measurement of orbital
precession and orbital decay (see also Sect.~\ref{sec:psrj}), triggered a number of
publications in the field of dynamical tides in wide eccentric binary systems
\cite[e.g.]{KG96,L97,KQ98}.  These papers considered some aspects of nonlinear coupling
of modes, effects of stellar rotation and enhanced dissipation due to differential
rotation by means of various approximations.  We will now investigate, by detailed
calculations, the effects of resonances with stellar oscillation modes (in a uniformly
rotating MS star) on the tidal evolution of eccentric massive binary systems and apply 
our results to \object{PSR J0045-7319}.

\section{Basic equations}
\label{sec:baseq}

We consider a binary system consisting of a $M_\mathrm{s} = 10\,\mathrm{M}_\odot$ main
sequence star and a $M_\mathrm{p} = 1.4\,\mathrm{M}_\odot$ compact (NS) companion in an
eccentric orbit with eccentricity $e$ and orbital period $P_\mathrm{orb}$.  Such a
system could be easily produced during the supernova explosion following a period of
mass transfer from the initially more massive primary to the rejuvenated MS object
\cite{HH72}.

The energy and angular momentum (magnitude) of the eccentric orbit is given by
\[ 
E_\mathrm{orb} =-GM_\mathrm{p} M_\mathrm{s}/2a \mbox{\ \ and\ \ } H_\mathrm{orb} =
\frac{M_\mathrm{p} M_\mathrm{s}} {M_\mathrm{p}+M_\mathrm{s}} a^2 \omega
\sqrt{1-e^2} 
\] 
where a is the semi-major axis and $\omega=2 \pi/P_\mathrm{orb}$
the mean angular velocity of the stars in their elliptic orbit.  For simplicity we 
assume the
stellar spin angular momentum vector $\mathbf{H}_\mathrm{s}$ to be aligned with the
orbital angular momentum vector.  In this paper, we will only consider secular tidal
changes in the magnitude of the orbital energy and angular momentum, and ignore
changes in the configuration of the orbit as given by precession of the axes (which
vanishes as the axes are assumed parallel) or advance of periastron (apsidal 
motion).

\subsection{The tidal potential}
\label{sec:tidpot}
\begin{figure}
  \includegraphics{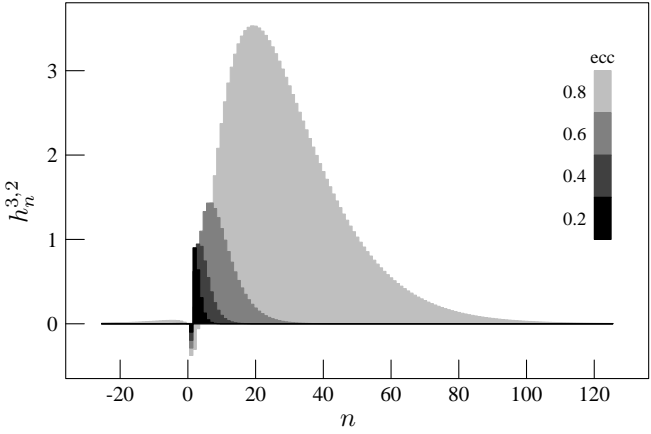}
  \caption[]{Hansen coefficients $h_n^{(l+1),m}$ with $l=m=2$ for eccentricities
    $e=0.2$, $0.4$, $0.6$ and $0.8$.}
  \label{fig:hansen}
\end{figure}

Due to the motion of the companion in its eccentric orbit, the 
$10\,\mathrm{M}_\odot$ MS star is
exposed to a changing external gravitational field.
We can facilitate the analysis by subdividing the forcing potential into its
harmonic components and evaluating the contribution of each term to the tidal
process separately.
Labelling the companion's coordinates relative to the star with a prime,
its perturbing potential is expanded as the real part of
\cite[e.g.]{MF53}
\begin{eqnarray}
  &&\Phi_\mathrm{T}(r,\vartheta,\varphi,t)= -\frac{G M_\mathrm{p}}{a} 
  \sum_{l=0}^{\infty}
  \sum_{m=0} ^l \epsilon_m \frac{\left( l-m\right)!}  {\left( l+m\right)!}
  \left(\frac{r}{a}\right)^l
  \nonumber
  \\
  &&\quad
  P^{m}_{l}(\cos \vartheta)
  \left(\frac{a}{r'}\right)^{l+1}  P^{m}_{l}(\cos
  \frac{\pi}{2})\, \mathrm{e}^{\mathrm{i}  m (\varphi' - \varphi)} 
  \nonumber
\end{eqnarray}
where $P^{m}_{l}(\cos \vartheta)$ denotes the associated Legendre polynomial and
$\epsilon_m=1$ for $m=0$ and 2 for $m>0$.  We assume $m\geq 0$ and characterise
retrograde relative orbital motion by a negative forcing frequency
$\bar{\sigma}=\sigma - m \Omega_\mathrm{s}$ in the stellar frame.  
Fourier expansion of the time dependent terms gives
\[
\left(\frac{a}{r'}\right)^{l+1} \mathrm{e}^{\mathrm{i} m (\varphi' -\varphi)} = 
\sum_{n=-\infty}^{\infty} h_n^{(l+1),m} \mathrm{e}^{\mathrm{i} (n M-m\varphi)}
\]
where $M = \omega t$ is the mean anomaly.
The Fourier coefficients (often called `Hansen coefficients' in this context) are 
given by
\[ 
h_n^{(l+1),m} = \frac{1}{2\pi} \int_{-\pi}^{\pi} \left(\frac{a}{r'}\right)^{l+1}
\mathrm{e}^{\mathrm{i}(m\varphi'-kM)} \,\mathrm{d} M.  
\]
The relative elliptic orbit can be expressed in the parameters $e$, $p$ and 
$\varphi'$ or in $e$, $a$ and the eccentric anomaly E:
\[  
r'=\frac{p}{1+e\cos \varphi'}=a(1-e\cos E).
\]
The eccentric anomaly $E$ is related to $M$ and $\varphi'$ through
\cite[e.g.]{BC61}
\[ 
E-e\sin E = M
\] 
and 
\[
\tan \left(\frac{\varphi'}{2}\right) = 
\sqrt{\frac{1+e}{1-e}} \tan \left( \frac{E}{2}\right).
\]
By scaling the Hansen coefficients as
\begin{equation}
  c_{\,n}^{lm} = \left(\frac{G M_\mathrm{p}}{a^{l+1}}\right) \epsilon_m 
  \frac{(l-m)!}{(l+m)!} 
  P_l^m\left(\cos \frac{\pi}{2}\right) h_n^{(l+1),m},
\end{equation}
the companion's tidal potential can be expressed as
\begin{eqnarray}
  &&  \Phi_\mathrm{T}(r,\vartheta,\varphi,t) = - \sum_{l=2}^{\infty}
  \sum_{m=0}^l 
  r^l    P_l^m(\cos \vartheta)
  \nonumber
  \\
  &&\quad
  \sum_{n=-\infty}^{\infty} c_{\,n}^{lm} \cos (n \omega t - m \varphi),
  \label{eq:tidpot}
\end{eqnarray}
i.e.\ each tidal spherical harmonic component $(l,m)$ can be expressed as a series of
harmonics $n$, representing circular orbits with angular frequency $n \omega$.  In this
paper we only consider the dominant tidal components with $l=2$ and $m=0$ or~$m=2$.
The $m=1$ contribution vanishes for the aligned case studied here.  Modes with $m=0$
have no $\varphi$ dependence, therefore no angular momentum can be transferred via
these oscillations, only energy exchange can take place.  For these axisymmetric modes,
the coefficients $c_{\,n}^{lm} = c_{\,-n}^{lm}$.  For $m=2$ the coefficients with $n<0$
are negligible.  In Fig.~\ref{fig:hansen} we plot the Hansen coefficients for $m=2$ as
a function of $n$ for a few different values of the eccentricity.  It is apparent that
as the eccentricity increases, the number of contributing harmonics increases greatly,
and also that the magnitude of the largest coefficients (which enters the expression
for the torque quadratically) increases by a significant factor due to the diminishing
orbital separation at periastron.

In our calculations we include $n$-values starting at $n=1$ and going up to
$n_\mathrm{max}=2 (m+l+1) f_\mathrm{per}$ to ensure that the companion's potential is
accurately reproduced.  For $m=0$ we take $c_{\,n}^{lm}$ to be the sum of the positive
and negative $n$ contribution.  Here, $f_\mathrm{per}=\sqrt{\frac{1+e}{(1-e)^3}}$ is 
the
ratio between the orbital frequency in periastron $\omega_\mathrm{per}$ and the mean orbital
frequency $\omega$.

\subsection{Orbital evolution by the tidal exchange of energy and angular momentum}
\label{sec:tidex}

Situated in an eccentric orbit with its companion, the $10\,\mathrm{M}_\odot$ star is
simultaneously forced at many harmonic frequencies $\bar{\sigma}_n=n \omega  - m
\Omega_\mathrm{s}$ (in the stellar frame), the magnitude of each term being
proportional to $c_{\,n}^{lm}$. The harmonic components with $n<n_0=m\,\mathrm{int}
(\Omega_\mathrm{s}/\omega)$ correspond to negative forcing frequencies which means
that the corresponding frequency in the inertial frame is smaller than the stellar
rotation frequency.  Hence in the stellar frame these components excite oscillations
which run backwards.  Excitation of such retrograde modes gives rise to spin-down of
the star, while excitation of prograde modes ($n>n_0$) has the opposite effect.
This opens the possibility of simultaneous tidal interaction with 
counteracting resonances with prograde (spin-up) and retrograde (spin-down)
oscillation modes.   By assuming the response of the star to the tidal
forcing can be approximated by a linear treatment the problem simplifies considerably
since we may apply the superposition principle and decompose the stellar response
into independent harmonic components.  We further assume the stellar response to each
harmonic has reached a steady state in which the tidal excitation is balanced by
radiative and viscous damping in the oscillating star.  These approximations may
break down during resonance passages although a steady state is usually a fairly
good approximation, see discussion in Sect.~\ref{sec:disc}. The steady state assumption
allows us to
calculate the stellar response as a strictly periodic phenomenon, i.e.\ we are not
forced to follow the stellar oscillations on a dynamical timescale, which, at the
current level and for the duration required to study secular evolution, would not be
possible with present day computer facilities.  The steady state tidal torque can be
determined by applying the implicit 2D code mentioned above whereby the periodic term
can be factored out.  In Paper~I we have shown that for each harmonic $(l,m,n)$ in
the forcing potential the work done by the tide (per time unit) on the star and the
associated rate of change of spin angular momentum $H_\mathrm{s}$ can be expressed as:
\begin{equation}
  \dot{E}_{n}^{lm} = \sigma_n \mathcal{T}_{n}^{lm}  \mbox{$\quad$ and $\quad$}
  \dot{H}_{n}^{lm} = m \mathcal{T}_{n}^{lm} \label{eq:tlm}
\end{equation}
whereby the torque integral is defined as
\begin{eqnarray}
  && 
  \mathcal{T}_{n}^{lm} = -\pi  c_{\,n}^{lm} \int_0^{R_s}\!\!\!\int_0^{\pi} 
  \mathrm{Im} 
  \left(\rho'(r,\vartheta)\right)\nonumber\\
  &&\quad
  P_l^m(\cos \vartheta )\, r^{l+2} \sin \vartheta \,\mathrm{d}\vartheta \,\mathrm{d}r
  \label{eq:torqint} 
\end{eqnarray}
where $\mathrm{Im}$ stands for imaginary part and $P_l^m(\cos \vartheta )$ is the 
associated Legendre polynomial of index $m$ and degree $l$. In steady state
$\dot{E}_{n}^{lm}$ equals the energy dissipation rate due to the $n^\mathrm{th}$ 
harmonic
of the tidal $(l,m)$ forcing.  For a given forcing frequency $\sigma_n= n \omega$ (in
the inertial frame) the tidal perturbation of the stellar mass density
$\rho'(r,\vartheta)$ occurring in the above integral follows from the tidal response
calculations in Paper~I, see current appendix.  Conservation of energy and angular
momentum then implies that the rate of change of orbital energy and angular momentum
follows by adding up the stellar rates of change in response to each harmonic term in
the tidal potential and then reversing the sign:
\begin{equation}
  \dot{E}_\mathrm{orb} = -\sum_{l,m} \sum_{n} \dot{E}_{n}^{lm} \mbox{$\quad$ and 
    $\quad$ }
  \dot{H}_\mathrm{orb} = -\sum_{l,m} \sum_{n} \dot{H}_{n}^{lm}.
  \label{eq:dEHdt}
\end{equation}
By expressing the orbital eccentricity $e$ in terms of the stellar masses and the 
orbital energy $E_\mathrm{orb}$ and angular momentum $H_\mathrm{orb}$ \cite[e.g.]{LL59}
\[ 
e=\sqrt{1 + \frac{2 M_\mathrm{tot}}{G^2 M_\mathrm{p}^3 M_\mathrm{s}^3} E_\mathrm{orb} 
  H_\mathrm{orb}^2} 
\]
with $M_\mathrm{tot}=M_\mathrm{p}+M_\mathrm{s}$, we can express the rate of change of 
the 
orbital eccentricity as
\begin{equation}
  \frac{\mathrm{d}e^2}{\mathrm{d}t}=\left(\frac{G M_\mathrm{p} M_\mathrm{s}}{2 
      a}\right)^{-1} \left[ 
    (1-e^2) \dot{E}_\mathrm{orb} - \omega \sqrt{1 -e^2} \dot{H}_\mathrm{orb} \right]
  \label{eq:deedt}
\end{equation}
We use this expression for the rate of change of the eccentricity to prevent numerical 
roundoff errors 
from producing negative eccentricities or generating eccentricity in a circular 
orbit. Finally, the rate of change of the semi-major axis follows from
\begin{equation}
  \frac{1}{a} \frac{\mathrm{d}a}{\mathrm{d}t}=\left(\frac{G M_\mathrm{p} 
      M_\mathrm{s}}{2 a}\right)^{-1} 
  \dot{E}_\mathrm{orb}. 
  \label{eq:daadt}
\end{equation}

By defining the eccentric orbit for given (fixed) stellar masses through $a$ and
$e$ the tidal evolution of the orbit can be followed by numerically integrating
Eqs.~(\ref{eq:deedt})--(\ref{eq:daadt}) whereby at every timestep the tidal
exchange rates $\dot{E}_\mathrm{orb}$ and $\dot{H}_\mathrm{orb}$ are determined
through Eqs.~(\ref{eq:tlm})--(\ref{eq:dEHdt}).

To investigate the resonant exchange of energy and angular momentum in detail, we
developed a routine that enables us to calculate this tidal evolution by interpolating
the tidal torque integral (\ref{eq:torqint}), as a function of forcing frequency
$\sigma_n$, stellar spin rate $\Omega_\mathrm{s}$ and evolutionary state of the MS star
(expressed in terms of the core hydrogen abundance $X_\mathrm{c}$) in the data
presented in Paper~I and Tables~\ref{tab:mmodes} and~\ref{tab:rmodes} (see appendix for
more details).

\subsection{Resonant exchange of energy and angular momentum}
\label{sec:resons}

We have seen that in an eccentric binary system the tidal forcing of the star by its
orbiting companion occurs in a non-harmonic time dependent manner. Decomposition of
the forcing potential into harmonic components introduces a range of frequencies at
which different harmonic modes of oscillation in the star are excited simultaneously.
These forcing frequencies change as orbital evolution progresses due to the fact that
the orbit may shrink or widen, and due to changes in the stellar rotation frequency.
Moreover, as the system evolves, the stellar oscillation spectrum itself changes due
to restructuring of the star in response to stellar evolution and due to spin-up or
spin-down of the star caused by tidal effects and the changing stellar moment of
inertia. These stellar changes result in shifting of the eigenfrequencies of the
stellar modes on the nuclear timescale of the star (which becomes relatively short,
with the possibility of rapid tidal evolution, near the end of core hydrogen burning
\cite{SP84}). Hence, as the orbit evolves, the forcing frequencies shift through the
stellar oscillation spectrum, picking up resonances in the star as eigen frequencies
are crossed.  During a resonance passage, the rate at which energy and angular
momentum exchange takes place between the star and the orbit (i.e.\ the companion)
can be many orders of magnitude larger than in the non-resonant case.  The total
amount of exchanged angular momentum is obtained by integrating the torque over time;
since the speed at which the forcing frequencies shift through the stellar
oscillation spectrum will generally be proportional to the torque, a relatively large
peak area in the torque--frequency graph does not guarantee a
large relative contribution to the tidal evolution process.  The angular momentum
exchange by a resonance passage of the $n^\mathrm{th}$ harmonic component crossing
resonance peak $k$ can be expressed as:
\begin{equation}
  (\Delta H)_\mathrm{res} = \int m\mathcal{T}_{n}^{lm} \,\mathrm{d}t 
  = \int\frac{m\mathcal{T}_{n}^{lm}}
  {\frac{\mathrm{d}(\bar{\sigma}_n-\bar{\sigma}_{0,k})}{\mathrm{d}t}}
  \,\mathrm{d}(\bar{\sigma}_n-\bar{\sigma}_{0,k})\label{eq:totH}
\end{equation}
with $\mathcal{T}_{n}^{lm}$ the torque integral~(\ref{eq:torqint}).  The rate at 
which a
certain harmonic component $n$ moves through the resonance with mode $k$ (at 
frequency $\bar{\sigma}_{0,k}$ in the stellar frame) is a  function
of the rate of change of the mean orbital and stellar angular speed and the
rate of stellar evolution (characterised by the rate of hydrogen depletion in the 
stellar core $\dot{X}_\mathrm{c}$):
\[
\frac{\mathrm{d}(\bar{\sigma}_n-\bar{\sigma}_{0,k})}{\mathrm{d}t}=
n\frac{\mathrm{d} \omega}{\mathrm{d} t}-
m\frac{\mathrm{d} \Omega_\mathrm{s}}{\mathrm{d} t}-
\frac{\partial \bar{\sigma}_{0,k}}{\partial X_\mathrm{c}}
\frac{\mathrm{d} X_\mathrm{c}}{\mathrm{d} t}-
\frac{\partial \bar{\sigma}_{0,k}}{\partial \Omega_\mathrm{s}}
\frac{\mathrm{d} \Omega_\mathrm{s}}{\mathrm{d} t}.
\]
After separating the dynamical and the stellar evolution component we obtain
\begin{eqnarray}
  &&
  \frac{\mathrm{d}(\bar{\sigma}_n-\bar{\sigma}_{0,k})}{\mathrm{d}t}=
  n\frac{\mathrm{d} \omega}{\mathrm{d} t}-
  \frac{1}{I_\mathrm{s}}
  \left(m+\frac{\partial \bar{\sigma}_{0,k}}{\partial \Omega_\mathrm{s}}\right)
  \frac{\mathrm{d}H_\mathrm{s}}{\mathrm{d}t}+\nonumber\\
  &&\quad
  \left[
    \frac{\Omega_\mathrm{s}}{I_\mathrm{s}}
    \left(m+\frac{\partial \bar{\sigma}_{0,k}}{\partial \Omega_\mathrm{s}}\right)
    \frac{\mathrm{d}I_\mathrm{s}}{\mathrm{d}X_\mathrm{c}}-
    \frac{\partial\bar{\sigma}_{0,k}}{\partial X_\mathrm{c}}
  \right]
  \frac{\mathrm{d}X_\mathrm{c}}{\mathrm{d}t},
  \label{eq:shift}
\end{eqnarray}
with $I_s$ the stellar moment of inertia.
We note that, while the second term on the right of Eq.~(\ref{eq:shift}) is
proportional to $\dot{H}_\mathrm{orb}$, the first term is, for given harmonic $n$,
fully determined by $\dot{E}_\mathrm{orb}$.  Of course the last term is determined by
stellar evolution.

\subsection{Resonance locking} 

When a harmonic term $n$ of the tidal potential comes into resonance with a stellar
oscillation mode the resulting exchange of energy and angular momentum between the MS
star and its companion usually leads to a rapid shift through the resonance, so that
the resonance condition is quickly lost and the tidal exchange is rather limited.
However, under special circumstances, the resonance condition may be sustained for a
relatively long period of time.  A necessary condition for such locking on to a
resonance is that the right hand side of Eq.~(\ref{eq:shift}) attains a small value
during resonance passage while, in order to drive a significant orbital evolution, the
acting torque should be large.  Substituting~(\ref{eq:tlm}) and introducing the
orbital moment of inertia $I_\mathrm{orb} = \frac{M_\mathrm{p}M_\mathrm{s}}
{M_\mathrm{p}+M_\mathrm{s}}a^2$, Eq.~(\ref{eq:shift}) can be expressed as:
\begin{equation}
  \frac{\mathrm{d}(\bar{\sigma}_n-\bar{\sigma}_{0,k})}{\mathrm{d}t}=  
  \epsilon \dot{X}_\mathrm{c} +
  \sum_{j} \zeta_{nj} \mathcal{T}_{j}^{lm} 
  \label{eq:mterms}
\end{equation}
where $l=m=2$ (the $m=0$ contribution is usually negligible). The dynamical factor is 
defined as
\[  
\zeta_{nj}=
\left[ \frac{3nj}{I_\mathrm{orb}}-
  \frac{m}{I_\mathrm{s}}\left(m+\frac{\partial \bar{\sigma}_{0,k}}{\partial 
      \Omega_\mathrm{s}}\right) \right] 
\]
while the stellar evolution factor is
\[ 
\epsilon=
\left[
  \frac{\Omega_\mathrm{s}}{I_\mathrm{s}}
  \left(m+\frac{\partial \bar{\sigma}_{0,k}}{\partial \Omega_\mathrm{s}}\right)
  \frac{\mathrm{d}I_\mathrm{s}}{\mathrm{d}X_\mathrm{c}}-
  \frac{\partial\bar{\sigma}_{0,k}}{\partial X_\mathrm{c}}
\right]. 
\]
Typically $\left| \frac{\partial \bar{\sigma}_{0,k}}{\partial \Omega_\mathrm{s}}
\right| \approx 0.5 < m$ for $m=2$ locking.  For the orbits we will consider
$I_\mathrm{orb}$ is typically two or three orders of magnitude larger than
$I_\mathrm{s}$.  For sufficiently large values of $n$ the diagonal factor
$\zeta=\zeta_{nn}$, which governs the frequency shift relative to the resonance peak
due to the action of the n$^\mathrm{th}$ harmonic itself, attains a small value.  A
harmonic term in the forcing potential which has a small self-shifting coefficient
$\zeta$ must be driven deep into a resonance before its influence on the frequency
shifting becomes prominent.  If its self-shifting tendency draws itself through the
resonance, the small value of $\zeta$ results in a relatively slow resonance
crossing with correspondingly enhanced exchange of energy and angular momentum.
However, if the self-shift works in the direction opposite to the resonance crossing,
the frequency shift (due to the combined action of the other harmonics and the effect
of stellar evolution) towards the resonance peak may be halted, resulting in a locked
condition.  This locked condition, whereby the self-shift balances the effect of all
other harmonics, may last for a long period, therefore the total effect on the orbit
can be significant.  The most efficient locking will occur when $\zeta \simeq 0$, i.e.\
for the tidal harmonic $n \simeq n_{\zeta} \simeq m \, \mbox{int}
\sqrt{\frac{I_\mathrm{orb}} {3 I_\mathrm{s}}}$, which depends on the system's orbital
period.  The locking condition depends on the distribution of the Hansen coefficients,
i.e.\ on the orbital eccentricity and on the location of the stellar oscillation
spectrum relative to the locked mode (on where the strongest modes are situated in
frequency space).

Consider the common case where the $n^\mathrm{th}$ harmonic $\bar{\sigma}_n$ approaches
a resonance $\bar{\sigma}_{0,k}$ from below, i.e.\
$\bar{\sigma}_n-\bar{\sigma}_{0,k}<0$ whereby $\zeta=\zeta_{nn}$ is slightly negative.
Assuming a MS star with 
prograde spin and $n_{\zeta}> n_0=m \,\mathrm{int} (\Omega_\mathrm{s}/\omega)$, where 
$n_0$ denotes the division between retrograde and prograde harmonics,
we can define
\[
S_1= \sum_{1\le j\le n_0} \zeta_{nj}
\mathcal{T}_{j}^{lm}  >0
\]
\[
S_2= \sum_{n_0<j<n} \zeta_{nj} 
\mathcal{T}_{j}^{lm}  <0 
\]
\[
S_3= \sum_{j>n} \zeta_{nj}    
\mathcal{T}_{j}^{lm} >0.
\]
Note that all harmonics with $j<n$ have $\zeta_{nj}<0$, while
all higher harmonics have $\zeta_{nj}>0$, while the torque 
integrals $\mathcal{T}_{j}^{lm}<0$ for $n<n_0$ and positive for  $n>n_0$.
Since $\zeta<0$ (but almost zero) the $n^\mathrm{th}$ harmonic tends to move away
from the resonance at $\bar{\sigma}_{0,k}$ and must be driven into resonance
by the combined action of the other harmonics (which will interact with other resonances)
and the weak (positive) effect of stellar evolution:
\[
-\zeta \mathcal{T}_{n}^{lm} < S_1 + S_2 + S_3 + \epsilon \dot{X}_\mathrm{c}.
\]
Locking requires that, when driven deep into resonance, the $n^\mathrm{th}$ harmonic
begins to dominate all other effects, so that its action can prevent resonance passage:
\begin{equation}
  -\zeta \max(\mathcal{T}_{n}^{lm}) > S_1 + S_2 + S_3 + \epsilon \dot{X}_\mathrm{c}.
  \label{eq:locking}
\end{equation}
The criterion is easily modified for a retrograde stellar spin or for $n> n_0$.
We will meet several cases of resonance locking in the numerical results discussed 
below.

\section{Results for moderately wide binary systems}
\label{sec:results1}

From now on we express all frequencies in units of $\Omega_\mathrm{c}$ and note that
we consider only the strongest tidal components of the companion's potential, i.e.\ 
the component for which $l=2$ with either $m=0$ or $m=2$.
Unless specifically stated, the initial evolution state of the MS star corresponds to
a central hydrogen abundance $X_\mathrm{c}=0.4$. 

Before we go on to the discussion of our numerical results, let us first express the
rate of change of the orbital eccentricity (\ref{eq:deedt}) due to the action of a
single resonance, by substituting the corresponding expressions (\ref{eq:tlm}) for
$\dot{E}^{lm}_n$ and $\dot{H}^{lm}_n$ as:  
\begin{equation} \frac{\mathrm{d}e^2}{\mathrm{d}t}=\left[
    \frac{\omega \sqrt{1-e^2}} {E_\mathrm{orb}}\right] \left( n \sqrt{1-e^2} -m \right)
  \mathcal{T}^{lm}_n \label{eq:deedt3} \end{equation} 
where the factor in square
brackets is always negative.  It follows that prograde harmonics ($n>n_0$) would 
require
$n<m/\sqrt{1-e^2}$ to yield a resonant increase of the orbital eccentricity.
Unless the eccentricity is unrealistically large these two conditions are mutually
exclusive, so that resonances with prograde modes tend to decrease the orbital
eccentricity.  On the other hand, retrograde modes which have $n<n_0$ and
$\mathcal{T}^{lm}_n<0$ require $n>m/\sqrt{1-e^2}$ to enlarge the orbital
eccentricity.  Depending on the stellar rotation rate these two conditions can
possibly be fulfilled, so that resonances with retrograde $m=2$ modes may enlarge the
orbital eccentricity.  Given the large number of harmonics that appear in orbits with
high eccentricities (of the order $10^2$ for $e \sim 0.8$), the orbital evolution of
such binary systems can be quite complicated.  However, before we study the high 
eccentricity case, we will
first consider the simple case of an almost circular orbit to gain some insight in
the effect of resonances.

\subsection{Case of a narrow, almost circular orbit}

\begin{figure*}[p]
  \includegraphics{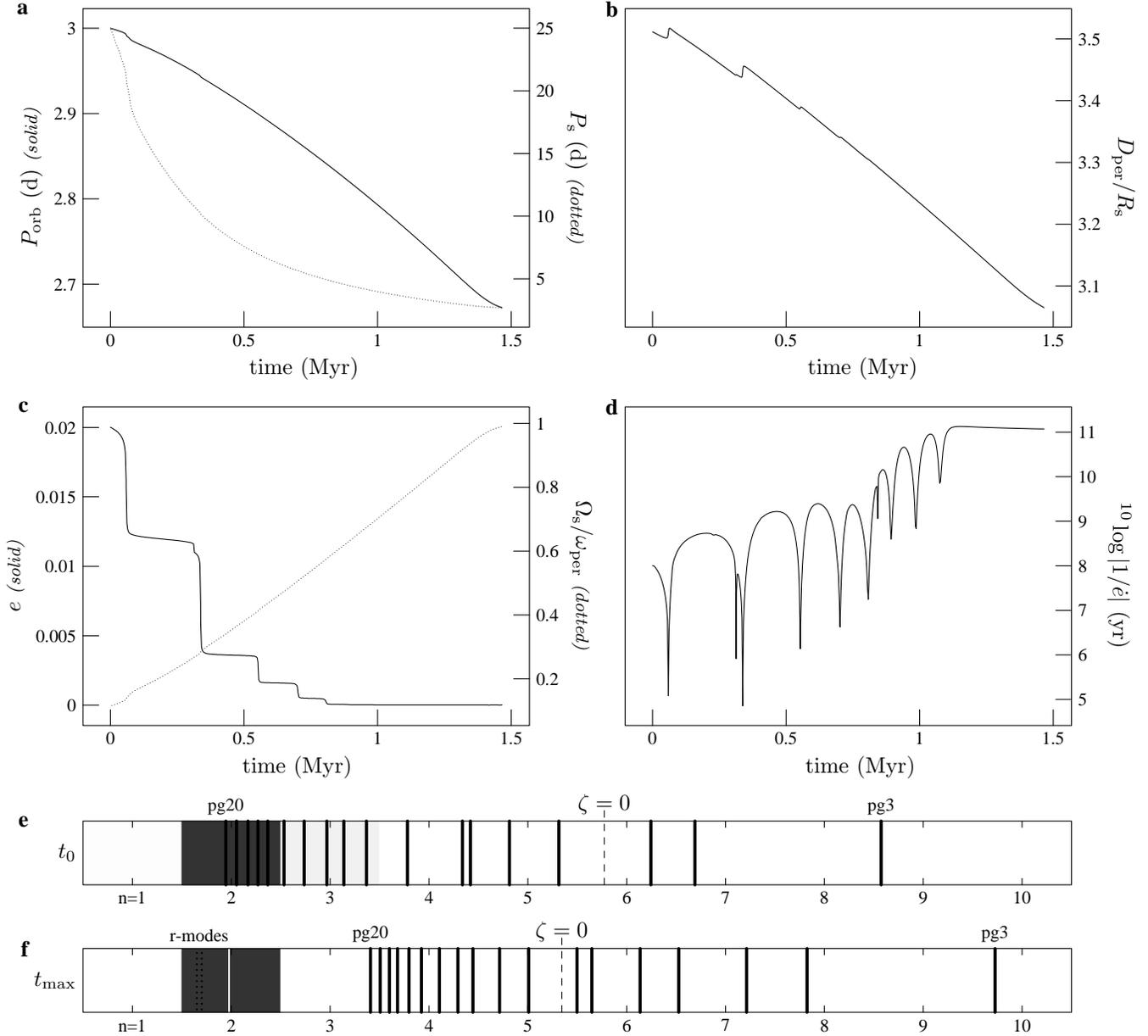}
  \caption[]{Orbital evolution of a weakly eccentric ($e=0.02$) binary system which
    initially, at $t=0$, harbours a slowly rotating MS star ($P_\mathrm{s}=25$~days)
    in an orbit with a period of
    three days.  During tidal evolution, resonant interaction through the $n=3$  
    harmonic 
    decreases the eccentricity to very low values ($e=0.00012$ at
    $t=1.5$~Myr), while non-resonant interaction via the $n=2$ component forces the 
    star
    into corotation.
    \textbf{a}~Orbital (solid, left ordinate) and stellar (dotted, right ordinate)
    rotation period,
    \textbf{b}~ratio of periastron distance to stellar radius,
    \textbf{c}~eccentricity (solid, left ordinate) and ratio of stellar
    rotation frequency to orbital frequency at periastron (dotted, right ordinate),
    \textbf{d}~timescale for the change of the eccentricity,
    \textbf{e}~and~\textbf{f}~schematic representation of the frequency distribution
    of forcing harmonics and stellar resonance frequencies at $t=0$ (panel~\textbf{e})
    and~$t=t_\mathrm{max}$ (panel~\textbf{f}). Here, frequency increases from left to 
    right,
    the contributing
    harmonic potentials have been labelled $n=1$ to~$10$~($=n_\mathrm{max}$) and
    the gray level of each cell centred on $n$
    reflects the magnitude of the corresponding Hansen coefficient $h_n^{3,2}$.
    Due to the very low eccentricities only the cell with $n=2$ (and for $t=0$, very 
    vaguely,
    $n=3$) is printed gray.
    The frequencies
    of the stellar g-modes are given by thick lines, while the location of the r-modes
    is in between the dotted lines. In panel~\textbf{e}, the r-modes are 
    outside
    the range of the forcing harmonics and do not show up. The dashed line gives the 
    frequency for which
    $\zeta\equiv\zeta_{nn}=0$ (see Eq.~(\ref{eq:mterms})). The white line (panel~\textbf{f})
    indicates the position where $\bar{\sigma}=0$.
    }
  \label{fig:evolecc}
\end{figure*}
First, we illustrate how the inclusion of resonant effects influences the tidal
evolution in case of an almost circular orbit, so that we have to deal with only few
harmonics. This we do by evolving a fairly narrow orbit ($P_\mathrm{orb}=3$ days)
with eccentricity $e=0.02$ and a slowly rotating stellar component ($P_\mathrm{s}=25$
days) until corotation and circularization is established, see Fig.~\ref{fig:evolecc}.
In Fig.~\ref{fig:evolecc}e we depict
the initial location (frequency) of the stellar $l=m=2$ modes in terms of the order
$n$ of the forcing harmonic and indicate which harmonics have the largest Hansen
coefficients for this orbit.  The $m=0$ modes play generally no prominent role in the
evolution.
For these low
eccentricities ($e^2 \ll 1$), we can use a low order expansion of the potential of
the orbiting companion in powers of the eccentricity to analyse our numerical
results.  To first order in $e$ the Hansen coefficients can be expressed as
$h_1^{3,0}=3e$, $h_1^{3,2}=-\frac{1}{2}e$, $h_2^{3,2}=1$ and
$h_3^{3,2}=\frac{7}{2}e$, yielding for the tidal potential
\begin{eqnarray}
  &&\Phi_\mathrm{T} 
  (r,\vartheta,\varphi,t)=-\left.\frac{GM_\mathrm{p}}{4a}\left(\frac{r}{a}\right)^2\right
  (
  - 6eP_2(\cos\vartheta) \cos\omega t \nonumber
  \\
  &&\quad
  - \frac{1}{2}e P_2^2(\cos\vartheta) \cos(\omega t-2\varphi)
  \nonumber
  \\
  &&\quad
  +  P_2^2(\cos\vartheta) \cos(2 \omega t-2\varphi)
  \nonumber
  \\
  &&\quad
  + \left. \frac{7}{2}e P_2^2(\cos\vartheta) \cos(3 \omega t-2\varphi)\right) + 
  O(e^2).
\end{eqnarray}
Note that the sign of the Hansen coefficients is irrelevant since the tidal torque 
integral is proportional to the square of the coefficients.
The (here prograde) $n=2$ harmonic is by far the strongest for low eccentricities.  
During orbital
evolution, its excitation frequency $\bar{\sigma}_2$ only crosses the resonance 
frequency
of the weak prograde g$^2_{20}$-mode (i.e.\ the g-mode with $l=2$ which has 20 radial
nodes). This happens at $t\approx 7.2\times 10^4$~yr, only 14~kyr after the
(intrinsically weaker) $n=3$ harmonic crosses the much stronger g$^2_{13}$ resonance 
peak.
Around this time, the orbital period shows a small dip, mainly due
to the $n=3$ resonance crossing.
Because the g$^2_{20}$ mode is the lowest frequency g-mode considered here, no
further resonance crossings occur with the $n=2$ harmonic during the remaining
orbital evolution. 
As the eccentricity drops during the orbital evolution, the $n=3$ Hansen coefficient is 
further
decreased and the following $n=3$ resonance passages induce little or no features in 
the
curve of the orbital
period. Due to the off-resonant $n=2$ forcing frequency the star steadily spins up 
towards
corotation.

For the $n^\mathrm{th}$ harmonic the time evolution of the 
eccentricity~(\ref{eq:deedt3}) can be 
expressed as
\begin{equation}
  \frac{\mathrm{d}e^2}{\mathrm{d}t} = 2 
  \frac{m-n}{H_\mathrm{orb}}\mathcal{T}_{n}^{lm}+O(e^2).\label{eq:deedtsme}
\end{equation}
Note that, initially, the star spins very sub-synchronously, so that the torque
acting on the star $m \mathcal{T}_{n}^{lm}>0$ for all $n$.  The $n=2$ harmonic of
the $m=2$ potential does not give rise to significant changes in the eccentricity of
the orbit because its circularizing effect vanishes to order $O(e^2)$.  It appears
that the potential term with $n=3$ is responsible for the strongest circularization
effect, while the $n=1$ component could potentially increase the orbital eccentricity
of the system.  However, whereas the frequency $\bar{\sigma}_1$ is so small that it
does not cross any prograde resonance during the evolution of the orbit
(although it does cross the r-modes around $t=0.8$~Myr), the $n=3$ forcing
frequency starts between the g-modes with $k=12$ and $k=13$, and passes through all
of the modes g$^2_{13}$ to g$^2_{20}$ as the evolution progresses.  Due to the small
ratio $(h_3^{3,2}/h_2^{3,2})^2=(\frac{7}{2}e)^2$ the total torque on the star is
dominated by the off-resonant $n=2$ contribution even when the $n=3$ forcing
frequency is close to an eigen-frequency of the star.  Therefore the rate
$\frac{\mathrm{d}\bar{\sigma}_3}{\mathrm{d} t}$ at which the $n=3$ frequency shifts
through the spectrum varies only little. The g-modes with thirteen to twenty radial
nodes roughly have $\Delta\bar{\sigma}\approx 10^{-3} \bar{\sigma}_0$. As only a few
percent of the resonance peak area for such a peak is further than 10 peak widths
away from the peak
location, while the peaks are typically $10^2$ peak widths apart, 
the tidal exchange rate due to the $n=3$ harmonic, and therefore the rate of
change of the orbital eccentricity, is not equally distributed in time.  Indeed, from
Fig.~\ref{fig:evolecc}c it is seen that only during a resonance passage the
eccentricity of the system is significantly reduced, while only small
eccentricity changes occur when $\bar{\sigma}_3$ is in between resonances.

\subsection{Example of resonance locking in a moderately eccentric ($e=0.25$)
  5 day orbit}\label{sec:lock} 
\begin{figure*}[p] 
  \includegraphics{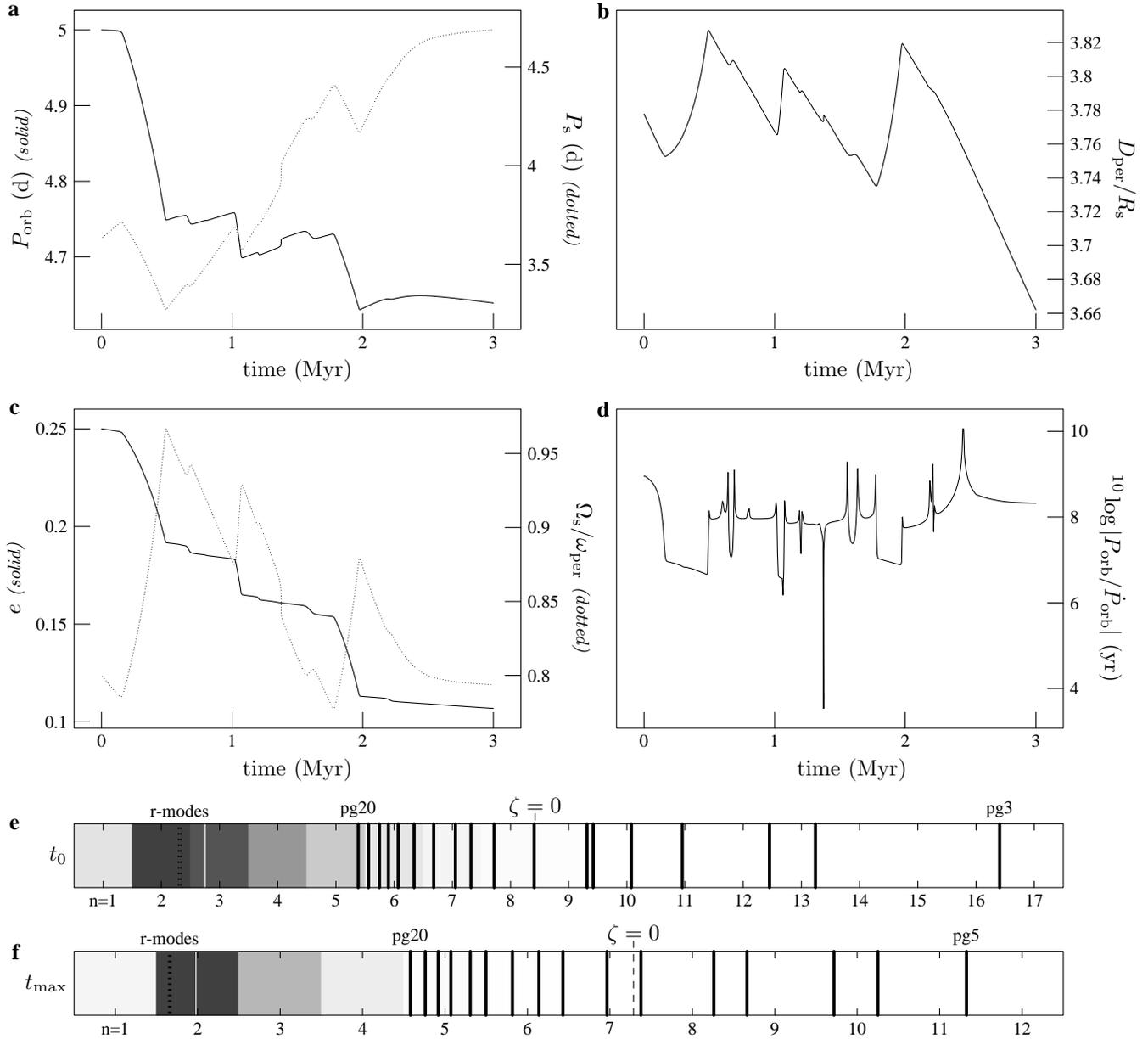} 
  \caption[]{Orbital evolution of a
    system with initial period $P_\mathrm{orb}=5$~days and eccentricity $e=0.25$.
    Initially the star rotates sub-synchronously 
    ($\Omega_\mathrm{s}/\omega_\mathrm{per}=
    0.8$), but resonance locking of the $n=7$ harmonic of the tidal forcing quickly
    forces the star to spin near (pseudo) corotation.  During the subsequent tidal
    evolution of the system, a few more cases of resonance locking (now with the $n=6$
    harmonic) occur, recognisable by the more or less square shape of the dips in panel
    \textbf{d}, which shows the timescale for the orbital decay.  
    \textbf{a}~Orbital and stellar rotation period, 
    \textbf{b}~ratio of periastron distance to stellar radius,
    \textbf{c}~eccentricity and ratio of stellar rotation frequency to orbital 
    frequency
    at periastron, 
    \textbf{d}~timescale for the change of the size of the orbit,
    \textbf{e}~and~\textbf{f}~schematic representation of the frequency distribution
    of forcing harmonics and stellar resonance frequencies at $t=0$ (panel~\textbf{e})
    and~$t=t_\mathrm{max}$ (panel~\textbf{f}).}
  \label{fig:evol4} 
\end{figure*}
Let us now consider a moderately eccentric binary system with initially $e=0.25$ and 
orbital
period $P_\mathrm{orb}=5$~days, which has a periastron separation of
$D_\mathrm{per}/R_\mathrm{s} \approx 3.8$.
The initial rotation speed of the MS star is adopted to be $0.8$~times the angular 
speed
of the orbiting companion in periastron, or $P_\mathrm{s}\approx 3.6$ days.
For this moderately eccentric orbit
the harmonics of the tidal forcing run from $n=1$ to $n_\mathrm{max}=17$ in our
calculation.  The maximum Hansen coefficient at $t=0$ corresponds to $n=2$:
$h_2=0.85$, while $h_n<\frac{1}{2}\, h_2$ for $n\ge 4$.  The orbital moment of inertia 
of the
five day orbit is $I_\mathrm{orb}= 9.1\times10^{57} $~g~cm$^2$, while the stellar
moment of inertia is $I_\mathrm{s}=1.6\times 10^{56}$~g~cm$^2\simeq
I_\mathrm{orb}/57$.  Since $\frac{\partial \bar{\sigma}_{0,13}}{\partial
  \Omega_\mathrm{s}}\simeq-0.15$ only, the factor $\zeta=\left[
  \frac{3n^2}{I_\mathrm{orb}}- \frac{2}{I_\mathrm{s}}\left(2+\frac{\partial
      \bar{\sigma}_{0,13}}{\partial \Omega_\mathrm{s}}\right) \right] $ which appears 
in
Eq.~(\ref{eq:mterms}) becomes positive for $n=9$, so that we expect locking to occur
(if it occurs at all) with the harmonic $n=7$ or~$8$, which appears indeed to be the
case.  Because $\bar{\sigma}_n$ and therefore $\mathcal{T}_{n}^{lm}$ is negative for
$n \leq n_0=2$ and positive for $n>2$, the locking conditions (\ref{eq:locking})
apply when we consider locking of $n=7$.  Because of the relatively narrow and
moderately eccentric orbit only a few low-order harmonics play a significant role in
the orbital evolution of this system and the sums $S_{i}$ in the locking conditions
consist now of single terms.  In Fig.~\ref{fig:evol4} we plot the orbital evolution
of this system.  Three phases of rapid tidal dissipation can be discerned during
which the orbital period and eccentricity decrease substantially.  During the first
phase of rapid evolution, around $t=0.2$ Myr, the initially sub-synchronously
spinning star ($\Omega_\mathrm{s}=0.8\, \omega_\mathrm{per}$) is spun up to almost
pseudo-corotation, while the orbital eccentricity decays from $e=0.25$ to $\approx
0.19$, in a few $10^5$ years.  This is a clear sign of resonance locking.

\begin{figure}[t] 
  \includegraphics{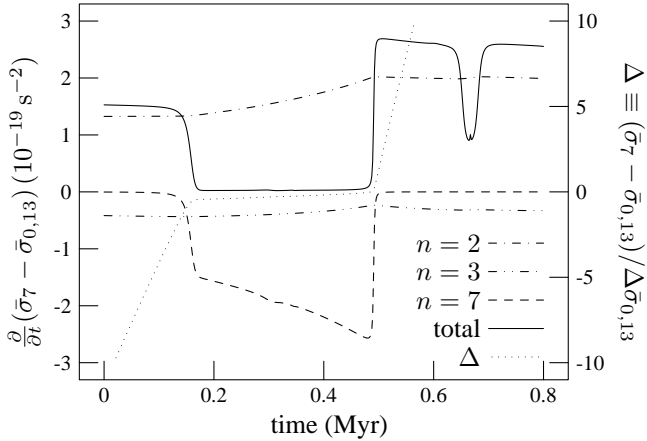} 
  \caption[]{Locking of the tidal harmonic $n=7$ onto the resonance with g$^2_{13}$. We 
    plot the dominant contributions to the (rate of) frequency shifting of the $n=7$ 
    harmonic relative to the resonance peak. The solid line gives the net rate of 
    shifting by summing over all harmonics and adding the almost constant positive 
    contribution of
    $\simeq 10^{-19}\mbox{ s} ^{-2}$ due to stellar evolution. The dotted line 
    represents 
    $\Delta=(\bar{\sigma}_7-\bar{\sigma}_{0,13})/\Delta \sigma$, i.e.\ the distance to 
    the 
    peak of the resonance in units of the FWHM of the resonance peak.} 
  \label{fig:lock} 
\end{figure}
However, at $t=0$ no mode is resonantly excited, and the forcing frequency 
$\bar{\sigma}_7$,
which is at the low frequency side of the prograde g$^2_{13}$ mode, drifts towards
the $k=13$ resonance peak mainly due to the combined effect of the star's 
evolution and driving by $n=2$.  This $n=2$ harmonic excites a non-resonant, heavily
damped low frequency (short wavelength) response at a frequency in between retrograde
g$^2_{20}$ and the r-modes.  The $n=2$ forcing thus induces a shift $\Delta
\bar{\sigma}_7$ which is insensitive to the exact value of $\bar{\sigma}_2$ and, for
a given magnitude of its Hansen coefficient, we can assume its effect to be
approximately constant.  Although the retrograde $n=2$ forcing term gives rise to the
strongest \emph{frequency shift}, its widening effect on the orbit is small compared
to the orbital decay caused by the prograde forcing with $n\geq 3$, especially by the
$n=7$ harmonic which approaches the resonance with g$^2_{13}$.  This relatively strong
mode is forced with high $n$ and thus efficient in bringing down the orbital
eccentricity and orbital period.  In Fig.~\ref{fig:lock} we zoom in on the g$^2_{13}$
resonance passage of the $n=7$ harmonic and plot the dominant contributions to the
frequency shift of this harmonic.

At about $t=0.1$~Myr, $\bar{\sigma}_7$ runs into the g$^2_{13}$ resonance peak, and due
to the negative sign of $\zeta_{7,7}$ the associated tidal torque tends to shift the
$n=7$ harmonic in the negative direction, back out of the resonance.  However, the
thermonuclearly driven expansion of the MS star counteracts this tendency, while the
retrograde $n=2$ harmonic pushes the $n=7$ harmonic even stronger into resonance.
Although the g$^2_{13}$ resonance gives an intrinsically much stronger response than
the non-resonant $n=2$ response, the high order $n=7$ forcing is significantly weaker
$\left((h_2/h_7)^2 \simeq 3.4\times10^2\right)$ so that this effect is moderated.  The
$n=7$ harmonic thus gets trapped by the combined action of stellar evolution and the
retrograde $n=2$ harmonic which drive the system into resonance against the resisting
self-shift of the locked harmonic.  Thereby the $n=7$ harmonic remains in close
resonance with g$^2_{13}$ during a few times $10^5$ years, see dotted line in
Fig.~\ref{fig:lock}.  Thanks to the relatively small value of $|\zeta_{7,7}|$, this
locking occurs for a relatively large torque.  Associated with the continuous resonant
tidal interaction we see a rapid decline of the orbital eccentricity.  This,
however, weakens the tidal forcing of the $n=7$ harmonic end strengthens that of the
$n=2$ harmonic.  Thereby the $n=7$ harmonic is driven deeper and deeper into resonance
with g$^2_{13}$ until it can no longer withstand the growing positive frequency shift
due the $n=2$ component and passes the top of the resonance.  Then the locking is
broken and the $n=7$ harmonic suddenly moves through and away from the resonance.  A
similar, but shorter, locking phase occurs at $t=1.067$ Myr between $n=7$ and
g$^2_{12}$.  The dominant $n=2$ harmonic crosses the r-modes at $t=1.375$ Myr which can
be observed in Fig.~\ref{fig:evol4}d as a single deep spike.  Around $t=1.9$ Myr, the
orbit now being narrower and $\zeta=0$ for $n\simeq$7--8, a new prolonged locking phase
sets in with $n=6$ and g$^2_{14}$.  Fig.~\ref{fig:evol4}c shows that significant decay
($\simeq 20 \%$) of the orbital eccentricity occurs when a harmonic locks on to a (not
particularly strong) resonance.  During these relatively short (of order $\simeq 10^5$
years) locking phases the sub-synchronous star spins up considerably.  However, the
retrograde $n=2$ forcing, in combination with the continuous stellar expansion of the
MS star brings the rotation rate down to small values. This continues until, near the 
end of the calculation, the stellar spin rate $\Omega_s$ drops below the mean orbital 
speed $\omega$ after which the $m=n=2$ harmonic becomes prograde. This brings the 
stellar spin down almost to a halt. 

\subsection{Tidal relaxation of a 10~day orbit with high ($e=0.7$) eccentricity.}
\label{sec:tenday}

\begin{figure*}[p]
  \includegraphics{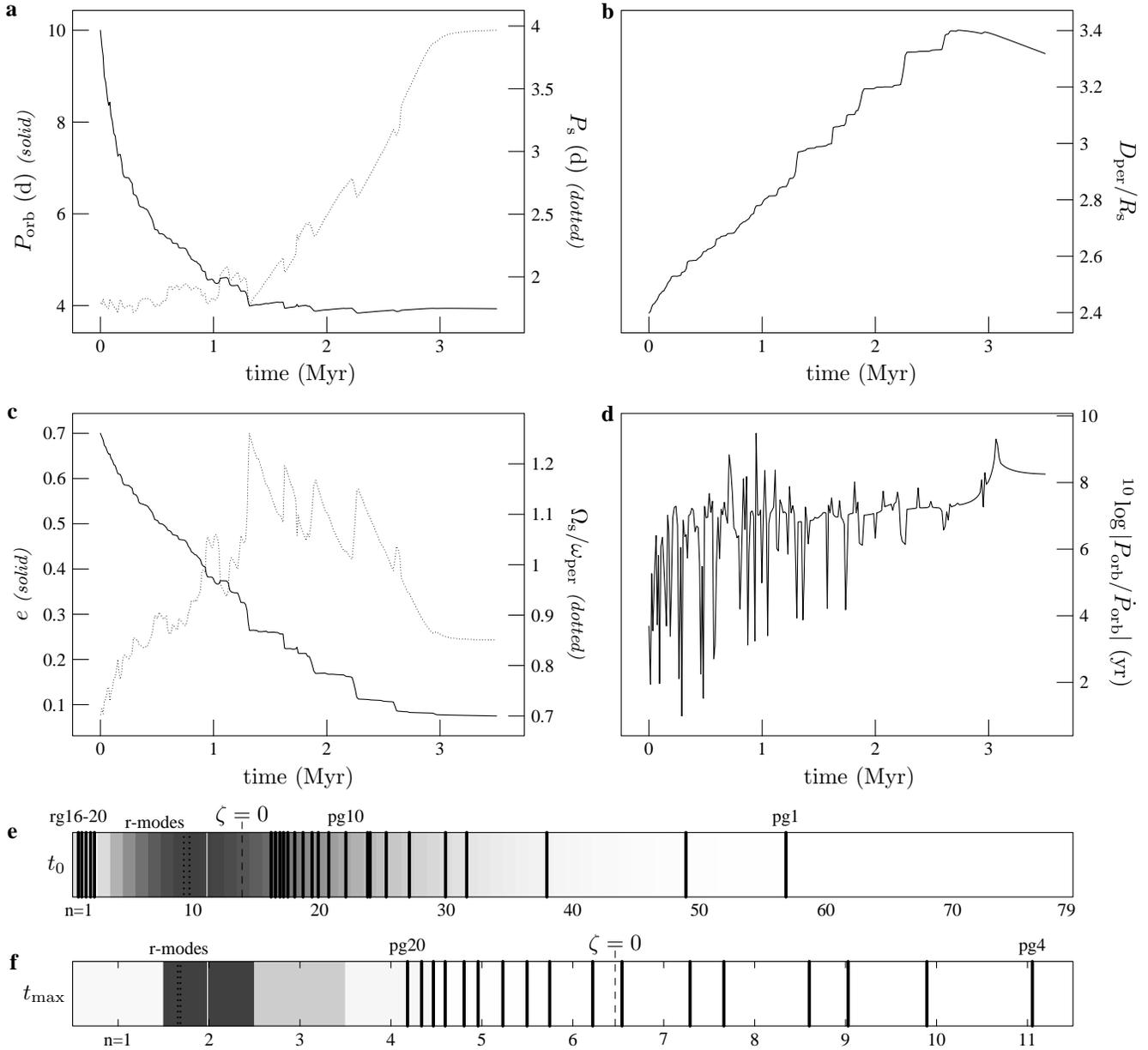}
  \caption[]{Orbital relaxation in a fairly wide orbit with high eccentricity.
    At $t=0$ the orbital period is $P_\mathrm{orb}=10$~days, the eccentricity
    is $e=0.7$, and the star has central hydrogen abundance $X_\mathrm{c}=0.40$.
    The star is initially rotating at seventy percent of the orbital angular speed at
    periastron.
    Within $3$~Myr the eccentricity is reduced to $\simeq 0.08$, and
    the orbit is shrunk to $\simeq 4$~days. At $t\simeq 3.1$~Myr the $n=2$ component
    in the forcing potential becomes prograde with respect to the stellar rotation.
    When its off-resonant spin-up effect comes into equilibrium with the spin-down 
    effect of the
    stellar expansion, the orbital evolution progresses on a nuclear timescale.
    At this time, the stellar rotation speed relaxed to $\simeq 0.85$ times 
    quasi-synchronisation.
    \textbf{a}~Orbital and stellar rotation period,
    \textbf{b}~ratio of periastron distance to stellar radius,
    \textbf{c}~eccentricity and ratio of stellar rotation frequency to orbital
    frequency at periastron,
    \textbf{d}~timescale for the change of the size of the orbit,
    \textbf{e}~and~\textbf{f}~schematic representation of the frequency distribution
    of forcing harmonics and stellar resonance frequencies at $t=0$ (panel~\textbf{e})
    and~$t=t_\mathrm{max}$ (panel~\textbf{f}).
    }
  \label{fig:tend}
\end{figure*}

Let us now illustrate the enhanced effectiveness of the tidal interaction in a highly
eccentric orbit by showing that a system which has a $10$~day orbit and $e=0.7$ will
evolve towards tidal relaxation on a timescale short compared to the MS lifetime of the
stellar component (which is approximately $2\times10^7$~yr).  Fig.~\ref{fig:tend} shows the
evolution of such an orbit; at $t=0$ we adopt a stellar rotation rate equal to $0.7$
times the orbital angular speed at periastron.  As is seen from panel~\ref{fig:tend}e,
strong forcing occurs in the r-mode range of the stellar oscillation spectrum.  Many
prograde g-modes, including fairly strong ones, are in the wing of the Hansen
distribution, creating many opportunities to transfer energy and angular momentum from
the orbit to the star.  Because of the location of $\zeta=0$, far from any resonance,
no locking is expected to occur during the initial stages of the orbital evolution.

Due to strong interaction with the stellar prograde g-modes the orbital period is
halved during the first one million years.  During this period the harmonics $n=9$
to~$5$ cross the r-modes, keeping the stellar rotation rate subsynchronous.

Between $t=1$~Myr and $t=3$~Myr the harmonic for which $\zeta\simeq 0$ shifts further
away from the r-modes, making the remaining r-mode crossings less effective, into the
region of the stellar oscillation spectrum where strong g-modes are to be found.  As a
result, resonance locking with prograde g-modes occurs, spinning the star up to
supersynchronous speed.  During this period the dominant retrograde forcing of the
$n=2$ harmonic, combined with the spin-down effect due to the stellar expansion, tends
to slow down the stellar rotation.  However, as long as resonance lockings of harmonics
with higher $n$-values occur, the star is kept at a supersynchronous rotation speed.
When around $t=3.1$~Myr the $m=n=2$ forcing frequency becomes prograde, the higher
harmonics are no longer driven through the stellar oscillation spectrum in a direction
opposite to their own shifting tendency, and equilibrium between two or more harmonics
that produces resonance locking is no longer possible.  At the end of the calculated
evolution, when the system is significantly closer to the tidal equilibrium situation,
the stellar spin-down due to stellar expansion and the spin-up due to the dominant
low-frequency $n=2$ forcing reaches approximate equilibrium, whereby the shift of the
forcing frequencies through the stellar oscillation spectrum progresses on the (long)
nuclear timescale of the MS star.  We expect full relaxation when the MS star
approaches the end of core hydrogen burning and the stellar evolution speeds up
considerably see \cite{SP84}.

\section{Results for wide binary systems: application to the binary pulsar \object{PSR 
    J0045-7319}}
\label{sec:psrj}
Let us now consider systems similar to the SMC binary radio pulsar \object{PSR
  J0045-7319}, which has an orbital period of 51.17 days and an eccentricity of 0.8080.
This binary system contains a B-star which, for an assumed $1.4\,\mathrm{M}_\odot$
neutron star, has a mass $(8.8 \pm 1.8)\mathrm{M}_\odot$.  From the observed rather
strong spin-orbit coupling the MS star is thought to be significantly deformed by
rotation, i.e.\ it is thought to rotate rapidly with its spin axis inclined with
respect to the orbital plane.  The periastron distance for
such a system is only $\simeq$ 4 stellar radii, and pulsar timing revealed that the
orbit decays on a timescale of $\frac{P_\mathrm{orb}}{\dot{P}_\mathrm{orb}}\simeq
-5\times10^5$~yr \cite[see]{BBSBK95,LBK95,KB96}.  Therefore, this system constitutes
an ideal test laboratory for the study of stellar tides and neutron star birth kicks.
According to Lai~\cite*{L96,L97}, the short timescale of orbital decay suggests a
significant retrograde stellar rotation with respect to the orbit.  Indeed, Brandt \&
Podsiadlowski~\cite*{BP95} show from Monte Carlo simulations that after a supernova
explosion, the spins of most stars in massive systems have large inclinations with
respect to their orbital axes, and a significant fraction of systems ($\sim 20$
percent) contain stars with retrograde spins.  Kumar \& Quataert \cite*{KQ97,KQ98}
argued that significant differential rotation, with a nearly synchronously rotating
surface layer around a rapidly spinning interior, may be required to explain the
observed timescale.

Although we cannot, at present, take into account the inclined orbit of \object{PSR
  J0045-7319} we can study some aspects of this complex system.  We will calculate the
evolution of an aligned system with approximately the same orbital period and
eccentricity, while adopting various
values for the initial spin rate of the MS star:  slightly super-synchronous rotation,
highly super-synchronous rotation and retrograde rotation of the MS star.
With `synchronous' rotation we mean here
of course a rotation rate equal to the orbital angular velocity of the companion at
periastron.  The initial eccentricity of $e\simeq 0.8$ implies that the periastron
frequency is $\omega_\mathrm{per}\simeq 15\,\omega$.  For such a highly eccentric
orbit the tidal forcing is decomposed into a large number (of order $10^2$) of
harmonics, so that resonance crossings become ubiquitous (e.g.\ see
Fig.~\ref{fig:evol1}d).  

\subsection{Case a:  slightly super-synchronous MS star} 

\begin{figure*}[p]
  \includegraphics{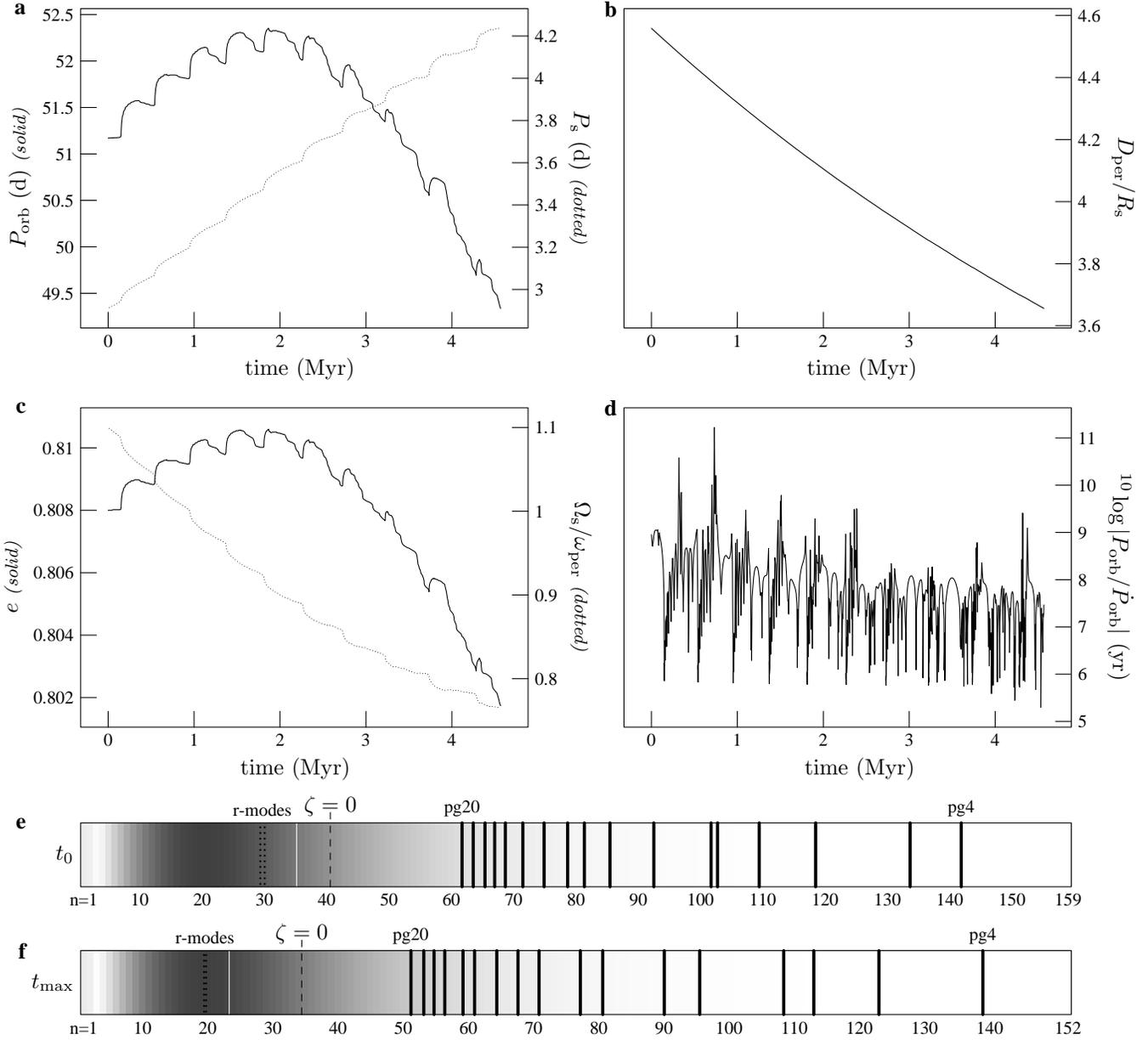}
  \caption[]{Orbital evolution of a very eccentric ($e=0.808$) binary system
    with a 51.17~day orbital period. 
    At $t=0$ the star rotates supersynchronously
    ($\Omega_\mathrm{s}/\omega_\mathrm{per}= 1.1$), but
    by the continuous stellar expansion and resonant interaction with the
    (retrograde) r-modes the originally super-synchronous star is spun down to $\simeq 
    0.8$ times periastron 
    frequency (dotted line panel \textbf{c}).
    Each time that one of the  tidal harmonics $n=29$ to $n=20$ runs through 
    the r-modes the orbital period and eccentricity increase. This corresponds to the
    repetitive bumps in panels \textbf{a} and \textbf{b} and in deep repetitive
    spikes in panel \textbf{d}. After $t 
    \simeq 2 $~Myr the prograde g-modes have shifted to lower frequencies (panel 
    \textbf{e}) and 
    become increasingly more prominent, so that the orbital separation and eccentricity 
    slowly begin to decay.
    \textbf{a}~Orbital and stellar rotation period,
    \textbf{b}~ratio of periastron distance to stellar radius,
    \textbf{c}~eccentricity and ratio of stellar rotation frequency to orbital
    frequency at periastron,
    \textbf{d}~timescale for the change of the size of the orbit,
    \textbf{e}~and~\textbf{f}~schematic representation of the frequency distribution
    of forcing harmonics and stellar resonance frequencies at $t=0$ (panel~\textbf{e})
    and~$t=t_\mathrm{max}$ (panel~\textbf{f}).
    }
  \label{fig:evol1}
\end{figure*}
Let us start with a $51.17$~day orbit, eccentricity $e=0.808$ and a
stellar rotation rate $\Omega_\mathrm{s}=1.1\, \omega_\mathrm{per}$, i.e.\ the MS star
is set to rotate 10\% faster than the largest angular speed of the companion.  This
corresponds to an initial stellar rotation period $P_\mathrm{s}\simeq 3 $ days.
In Fig.~\ref{fig:evol1}e it can be seen that the retrograde g-modes ($k\le 20$) fall outside the 
tidal window.
The dominant harmonics lie between $9 < n< 43$, all other harmonics have Hansen
coefficients at least 50\% weaker.  The r-modes all cluster around harmonic $n=29$,
near the Hansen-peak and are thus expected to be prominent in the following tidal
evolution.  The prograde g-modes ($n_0=35$) from g$_{20}$ to g$_{4}$ can be excited,
although they cover the range $n>62$ and thus lie in the weak wing of the Hansen
curve, so that the resonances will be weak.  The zero point of
$\zeta$ depends, for given $m=2$, only weakly on $k$ and
corresponds to $n=n_{\zeta}\simeq 40$.  Because there is no stellar mode which can be
excited by this harmonic, we do not expect resonance locking to occur.
Even when the stellar spin rate is decreased to zero the harmonics near $\zeta=0$
can excite only high radial order g-modes ($k\simeq 18$) which seem too weak to
induce efficient locking.

Let us now turn to the numerical results depicted in Fig.~\ref{fig:evol1}.
As expected, very soon the $n=29$ harmonic of the $l=m=2$ tidal forcing comes into
resonance with the fundamental r-mode r$^3_0$.  After running through the whole range
of r-modes (we consider $k=0$ to $k=10$ in our calculations, see Table~B2 in Paper~I)
the same process repeats itself with the $n=28$ harmonic, etc.\ until finally, at
$t\simeq 4.3\times 10^6$~yr, the $n=20$ harmonic runs through the r-mode range.  Each
time a harmonic runs through the r-mode range a new bump appears in the curve plotted
in Figs.~\ref{fig:evol1}a and~b because the r-modes transfer energy and angular
momentum from the rapidly spinning star into the orbit in such a way that the
semi-major axis and the eccentricity tend to increase.  Eq.~(\ref{eq:deedt3}) shows
that for the initial eccentricity $e=0.81$ the harmonics with $5<n<n_0=35$ will
increase the orbital eccentricity, and the r-modes fall in this range.  The
effect of the evolutionary spin down of the star and of these repeated 
r-mode passages is that the star rapidly spins down to
a sub-synchronous speed.  The repetitive crossings of r-modes can also be seen in
Fig.~\ref{fig:evol1}d.  In between r-mode crossings the system now and then crosses a
prograde g-mode ($n\simeq 70$--$80$ pass through g$_{10}$ to g$_{14}$)
giving rise typically to orbital $|P/\dot{P}|\simeq 10^6$ years. Due to the stellar 
spin down the prograde g-modes shift to lower frequencies so that stronger harmonics 
can excite them. This effect, in combination with relatively strong interaction with 
$m=0$ g-modes, becomes dominant after about $t \simeq 2$ Myr whereby the orbital 
separation and eccentricity begin to decay (slowly). After 4 million
years, the stellar rotation rate is reduced to approximately eighty percent of the
pseudo-corotation rate ($P_\mathrm{orb}\simeq 4.2$ days) while the eccentricity,
after an initial rise, is diminished by a small amount. We note that the effect of 
stellar expansion and  despinning r-modes prevent the MS star from attaining 
approximate corotation in periastron, even while the periastron separation is quite 
small in this system.

\subsection{Case b: highly super-synchronous MS star} 

\begin{figure*}[p]
  \includegraphics{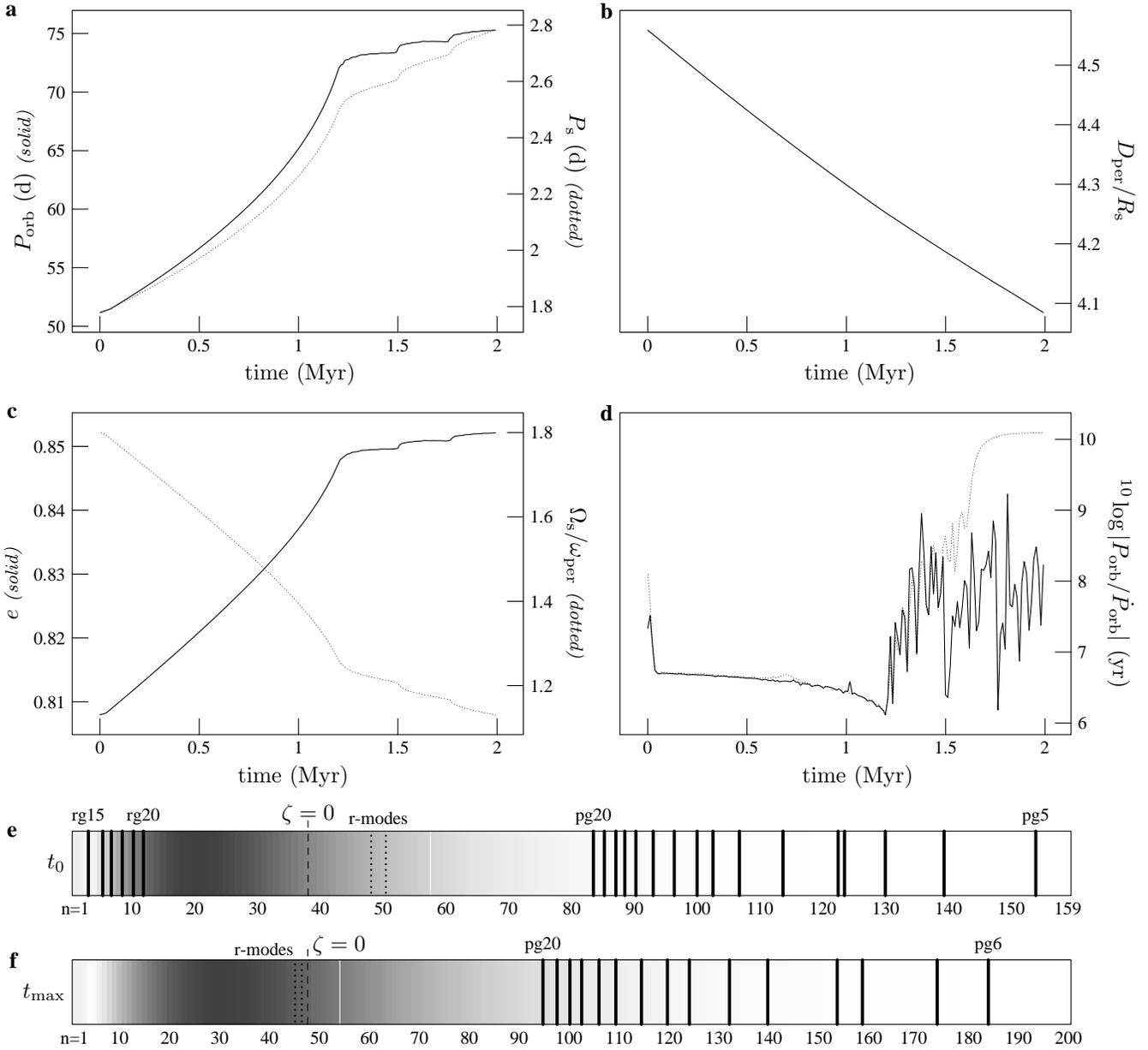}
  \caption[]
  {Orbital evolution of a very eccentric binary system. At $t=0$ the system's
    parameters $P_\mathrm{orb}=51.17$~days and $e=0.808$
    are the same as for the orbit in Fig.~\ref{fig:evol1}, but
    the stellar rotation speed in this case is higher; $1.8$ times the companion's
    angular speed at periastron.
    Very soon the $n=48$ harmonic in the potential runs into the fundamental r-mode
    and is locked near the resonance frequency for a period of more than one million
    years. During this period the star is spun down from $1.8$ times periastron
    speed to less than $1.3$ times periastron speed, while the orbital period
    is increased to approximately $72$ days. During the locking the orbital 
    eccentricity
    increases to almost $0.85$.
    \textbf{a}~Orbital and stellar rotation period,
    \textbf{b}~ratio of periastron distance to stellar radius,
    \textbf{c}~eccentricity and ratio of stellar rotation frequency to orbital
    frequency at periastron,
    \textbf{d}~timescale for the change of the size of the orbit,
    \textbf{e}~and~\textbf{f}~schematic representation of the frequency distribution
    of forcing harmonics and stellar resonance frequencies at $t=0$ (panel~\textbf{e})
    and~$t=t_\mathrm{max}$ (panel~\textbf{f}).
    }
  \label{fig:evol2}
\end{figure*}
\begin{figure}[t]
  \includegraphics{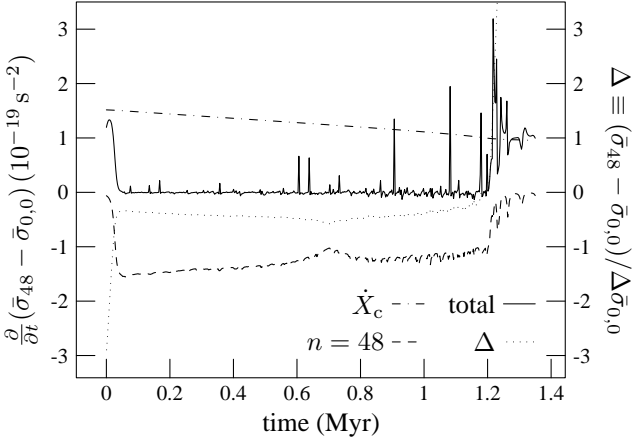}
  \caption[]{Locking of the tidal harmonic $n=48$ onto the resonance with r$^3_{0}$. We 
    plot the rate of frequency shifting of the $n=48$ 
    harmonic relative to the resonance peak. The contribution due to stellar
    evolution is plotted dot-dashed, the contribution of the $n=48$ excitation itself 
    is dashed.
    The solid line gives the net rate of 
    shifting by summing over all harmonics and adding the contribution due to stellar
    evolution;
    the dotted line represents 
    $\Delta=(\bar{\sigma}_{48}-\bar{\sigma}_{0,0})/\Delta \sigma$, i.e.\ the distance 
    to the 
    peak of the resonance in units of the FWHM of the resonance peak.
    }
  \label{fig:rrshift}
\end{figure}
In this case we adopt the orbital parameters of case a, but the star is initially set
to rotate at a higher rate: $\Omega_\mathrm{s}/\omega_\mathrm{per}= 1.8$.
We still consider the case of stellar rotation in the same direction as the orbital
motion of the companion.
Fig.~\ref{fig:evol2} gives the resulting orbital evolution, with the addition
of a dotted line in panel~\ref{fig:evol2}d which gives the tidal timescale in case
only forcing of the $n=48$ ($m=2$) harmonic would be taken into account.
During some $1.1$~Myr the tidal timescale, and therefore the acting torque, is 
dominated
by the single contribution of this $n=48$ harmonic, indicating resonance locking.

Fig.~\ref{fig:rrshift} displays the rate of frequency shifting of the $n=48$ forcing
frequency relative to the r$^3_{0}$ resonance frequency.  On average, the net rate of
shifting is kept close to zero due to cancellation of the stellar evolution (positive)
and $n=48$ (negative) contributions which are in equilibrium for some $1.1$~Myr.  The
resonance passages of intrinsically strong (low radial order) prograde g-modes which
occur (see Figs.~\ref{fig:evol2}e and~\ref{fig:evol2}f) induce only minor changes to
the orbit and to the locking condition, due to their small corresponding Hansen
coefficients, while passages of relatively strong harmonics through the retrograde
g-mode resonances are unimportant because the modes which are crossed have many radial
nodes and are thus intrinsically weak through strong damping. Although some of these
resonance crossings do show up as spikes in the solid curve in Fig.~\ref{fig:rrshift},
their duration is too short to significantly influence the resonance locking; many more
of these spikes are not plotted due to graphical limitations.  Only the passage of
$n=49$ through r$^3_6$ around $t=0.7$~Myr is able to briefly push the locked harmonic
further from resonance.  During evolution, $\zeta_{48,48}$ is positive, but its value
decreases as the orbit becomes wider.  The ability of the locked mode to resist being
pushed through resonance therefore weakens, and a larger torque is required to maintain
the equilibrium, causing the harmonic forcing to drift closer towards the resonance
peak.  Eventually, the resonance is crossed and the locking is ended, which happens
around $t=1.2$~Myr. After this happens the $\bar{\sigma}_{48}$ forcing frequency drifts
through the higher radial order r-modes, which have lower peak values and are unable
to restore the locking condition. The $n=47$ harmonic runs into the stellar r$^3_{0}$ mode
before the $n=48$ frequency reaches the last r-modes, and from this moment on the timescale
(panel~\ref{fig:evol2}d) is no longer dominated by the action of the previously locked
harmonic, but instead shows a sequence of dips below the dotted line which mark the
r-mode resonance crossings of $n=47$. Somewhat later ($t\simeq1.75$~Myr), the next harmonic
with $n=46$ reaches the r-modes, causing a similar sequence of dips in the timescale curve.

\subsection{Case c: a retrograde spinning MS star} 

\begin{figure*}[p]
  \includegraphics{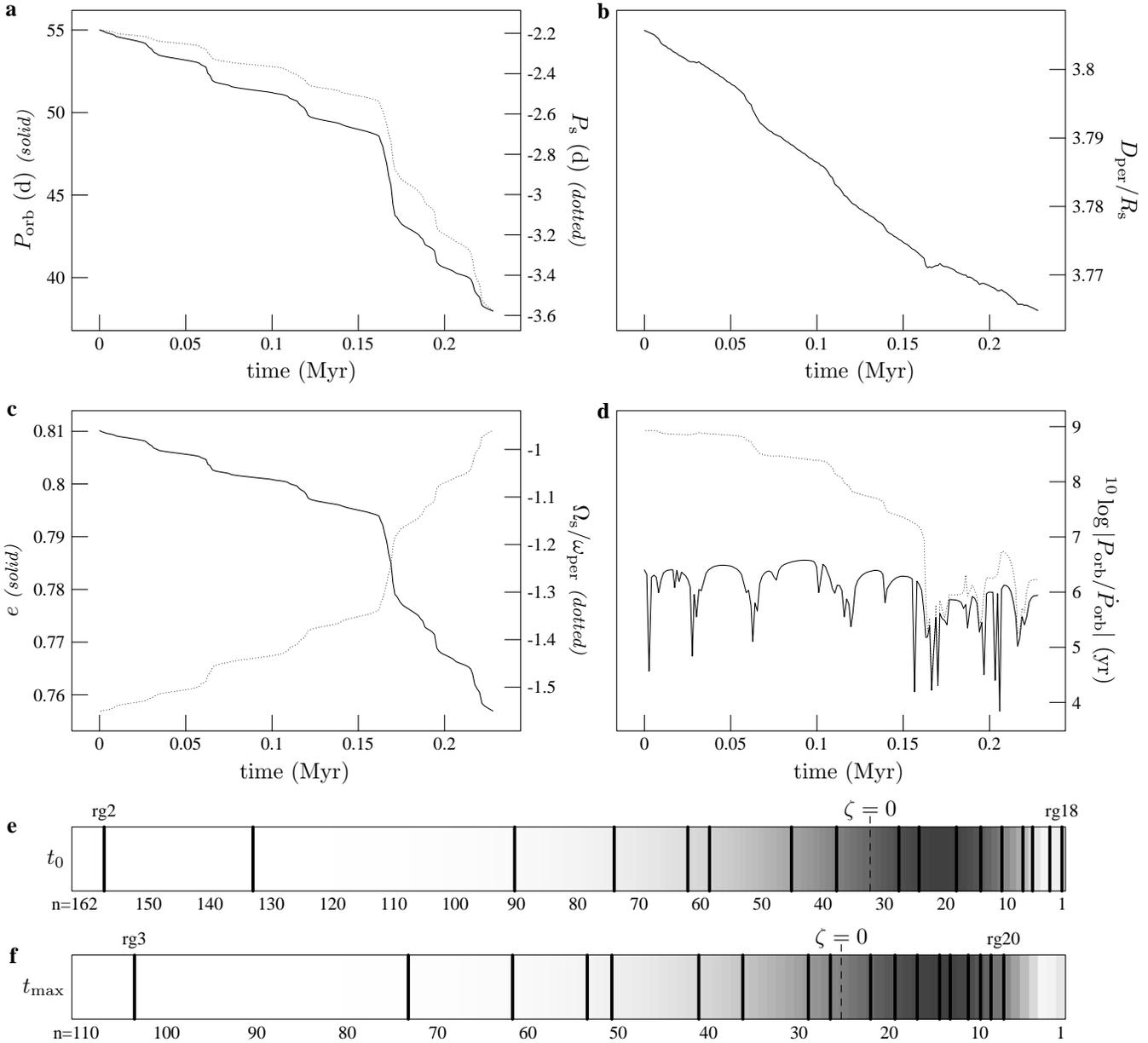}
  \caption[]{Orbital evolution of a system mimicking \object{PSR J0045-7319}.
    At $t=0$ the orbital period is $P_\mathrm{orb}=55$~days, the eccentricity
    is $e=0.81$, and the star has central hydrogen abundance $X_\mathrm{c}=0.21$
    so that
    the periastron distance is initially approximately $3.8$ stellar radii. The star 
    initially has a 
    retrograde
    rotation frequency of $\Omega_\mathrm{s}\simeq 0.3\,\Omega_\mathrm{c}$.
    The large number of contributing tidal harmonics in the early stages of
    the orbital evolution frequently cause the timescale for orbital evolution 
    (panel~\textbf{d})
    to reach values below $10^5$~yr. A total of approximately 1000 resonance 
    passages occur during
    the evolution shown, which is terminated as $X_\mathrm{c}$ reaches $0.2$.
    \textbf{a}~Orbital and stellar rotation period,
    \textbf{b}~ratio of periastron distance to stellar radius,
    \textbf{c}~eccentricity and ratio of stellar rotation frequency to orbital
    frequency at periastron,
    \textbf{d}~timescale for the change of the size of the orbit,
    \textbf{e}~and~\textbf{f}~schematic representation of the frequency distribution
    of forcing harmonics and stellar resonance frequencies at $t=0$ (panel~\textbf{e})
    and~$t=t_\mathrm{max}$ (panel~\textbf{f}).
    }
  \label{fig:psr2}
\end{figure*}
\begin{figure*}
  \includegraphics{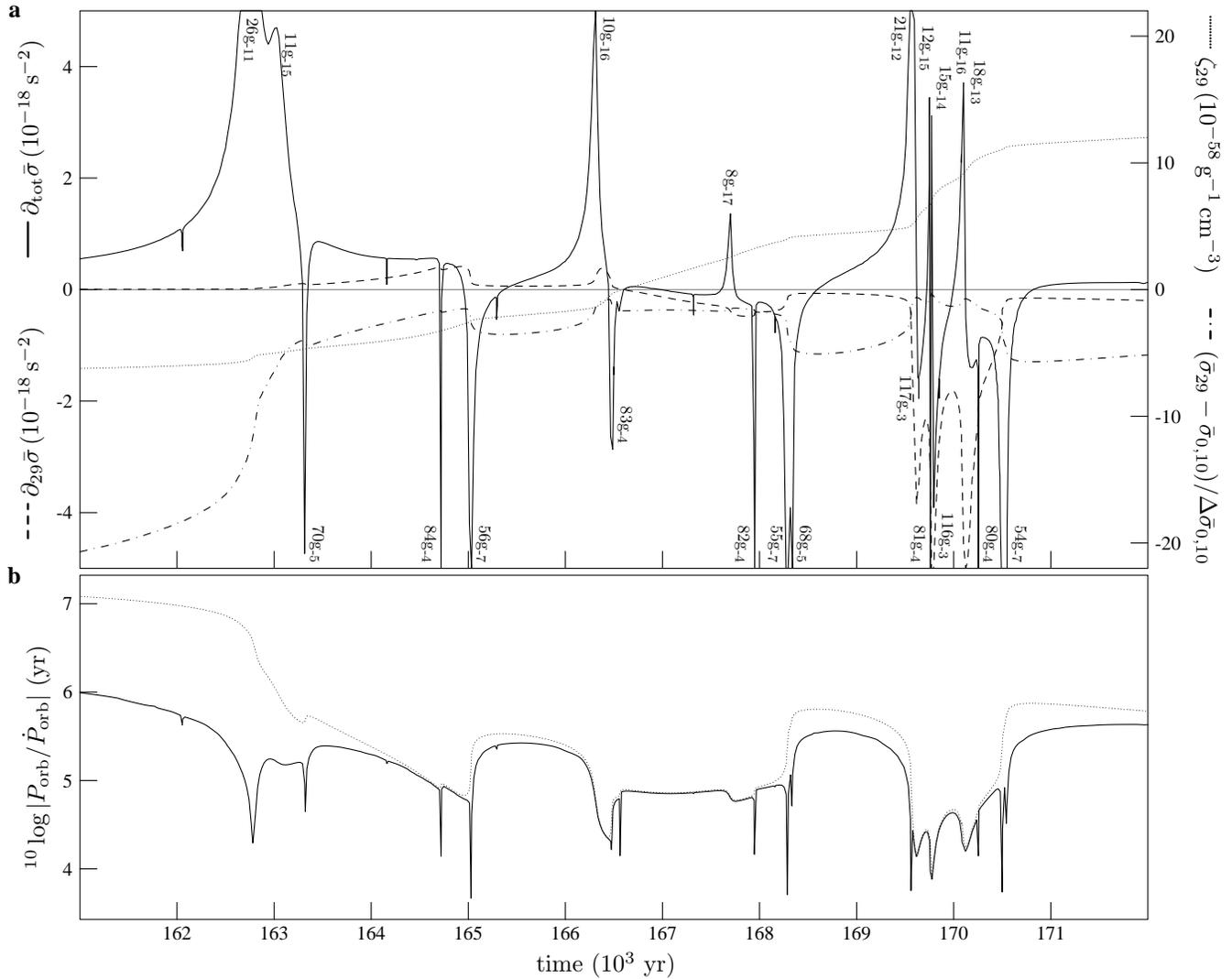}
  \caption[]
  {Resonance locking of $n=29$ forcing harmonic on the retrograde g$^2_{-10}$-mode.
    \textbf{a}~\emph{Left ordinate:} net rate of relative ($\bar{\sigma}_{29}-
    \bar{\sigma}_{0,10}$) frequency
    shifting due to the combined effect of all tidal harmonics and
    stellar evolution (solid line) and the shifting rate due
    to excitation of the $n=29$ term alone (dashed line). The nature of the resonance 
    crossings is indicated along the solid line as $n$g$_{-k}$, where $n$ gives the forcing 
    harmonic and $k$ the number of radial nodes of the excited oscillation.
    \emph{Right ordinate:} $\zeta_{29}$ (dotted line) and the distance between
    the $n=29$ forcing frequency and the $k=10$ retrograde g-mode resonance frequency.
    \textbf{b}~Detail of panel~\ref{fig:psr2}d; characteristic timescale of orbital evolution 
    (solid
    line) and characteristic timescale due to $n=29$ forcing alone (dotted line).
    }
  \label{fig:rgshift}
\end{figure*}
We will now consider the case where the star is spinning in the sense opposite to
the orbital revolution of its compact companion, at thirty percent 
of its surface break-up speed. Furthermore we will adopt an initial orbital period 
$P_\mathrm{orb}=55$ days, an eccentricity $e=0.81$  and a core hydrogen abundance
$X_c=0.21$, thus a fairly 
evolved MS star. For this initial configuration the tidal window is completely filled 
up with retrograde (in stellar frame) g-modes, see Fig.~\ref{fig:psr2}e. The peak of 
the Hansen distribution lies around harmonic $n=21$, while the $\zeta$-factor 
vanishes for $n \simeq 32$ which is near the resonance with $g^2_{-10}$. Thus under 
favourable circumstances this resonance may get locked, as will indeed be the case. 
Following the system until the core hydrogen content has been diminished to twenty
percent, the orbital/spin evolution proceeds as depicted in Fig.~\ref{fig:psr2}.

At about $t=0.162$ Myr the $n=29$ harmonic approaches the resonance with g$^2_{-10}$,
see Fig.~\ref{fig:psr2}a.  The factor $\zeta_{29,29}$, or $\zeta_{29}$ for short as
defined in~(\ref{eq:mterms}), being small from the very beginning, approaches zero and
changes sign at $t \simeq 0.167$ Myr.  From this moment on, the self-shifting direction
of the $n=29$ excitation is counter to the direction it has been shifting in, and it
stops itself from passing the $g^2_{-10}$ resonance frequency.  Also, since
$\zeta_{29}$ is so close to zero, the torque associated with the resonant
interaction hardly leads to a shift of the forcing frequency $\bar{\sigma}_{29}$ and
the locking can be very efficient.  It depends on the action of the remaining harmonics
whether the locking will last.  The lower order harmonics with $n<29$ tend to push the
$n=29$ forcing in the positive direction, i.e.\  deeper into resonance while the
harmonics with $n>29$ tend to shift it back away from the resonance.  Although the
latter harmonics can excite strong low radial order g-modes, the peak of the Hansen
distribution occurs for $n\simeq 20$, so that the low order harmonics dominate and keep
the $n=29$ forcing close to resonance, see Fig.~\ref{fig:rgshift}.  The dotted curve in
Fig.~\ref{fig:psr2}d and Fig.~\ref{fig:rgshift}b represents the sole contribution of
the $n=29$ harmonic to the decay time of the orbital period.  It can be seen that from
$t\simeq 0.164$ Myr to $t\simeq 0.171$ Myr most of the orbital decay is caused by the
resonant interaction between $n=29$ and g$^2_{-10}$.  During some 13~kyr the orbital
decay time is between $10^5$ and $5 \times 10^5$ years, which is the observationally
estimated decay time in \object{PSR J0045-7319}.  Later in the evolution the $n=29$
harmonic causes more phases with a similar short decay time while exciting g$^2_{-10}$,
but the $\bar{\sigma}_{0,-10}$ resonance frequency is never crossed.

In our calculation with retrograde but aligned spin locking occurs for conditions slightly off 
the observed $P_\mathrm{orb}$ and $e$.  But note that the system parameters for which locking
occurs depend on the details of the stellar input model and on the procedure for
interpolating between the two evolution states $X_c=0.4$ and $X_c=0.2$. We conclude 
that resonance locking might be responsible for the observed fast orbital decay in this 
system.

\section{Discussion}
\label{sec:disc}

We have studied the tidal evolution of eccentric early type binary systems by
decomposing the time dependent tidal potential in a series of harmonics and by calculating
the combined linear response of the uniformly rotating MS star with help of 2D-implicit
code.  We assumed the tidal oscillations to be in a steady state, whereby the tidal
excitation is balanced by radiative damping.  This allowed us to follow the orbital and
spin evolution of the binary system in detail.  However, during resonance crossings the
assumptions of steady state and linearity may break down.  Resonance peaks may
sometimes be crossed on timescales short compared to their corresponding damping
timescale $\tau_\mathrm{d}$.  As build-up of the full steady state response will take a
number of damping times one may question the validity of the steady state approximation
under these circumstances.  In reality the resonance crossing will be slower, as the
tidal response is not yet fully developed.  However, the tidal evolution is not
expected to be much different from our steady state results because the real evolution
timescale is still likely to be significantly shorter than the timescales associated
with the other (non-resonant) harmonics.

Non-linear effects can become significant during a resonance crossing with a weakly
damped mode and may give rise either to enhanced or reduced tidal evolution, depending
on circumstances.  But note that non-linear effects may often be of less importance
because strong tidal
interaction generally implies rapid orbital and spin evolution away from the
resonance, so that the oscillation amplitude is already limited by linear effects.
Non-linear effects may influence resonance locking, whereby a harmonic remains near a
resonance for a prolonged period of time.  Non-linear damping might lower the
resonance peak height and might thus diminish to some extent the duration of locking.

On the other hand, we have ignored the effect of resonances with $l>2$.  In a rotating
star the dominant $l=2$ tidal component can excite (through the Coriolis force) also
$l=$4,\,6,\,8 etc.\  modes, see Paper~I.  These higher spherical degree resonances will
add additional resonance crossings during tidal evolution although for moderate stellar
spin rates the peak area of these resonances is much less than that of the $l=2$ peaks.
However, since peak area is not important for resonance locking, resonance locking
may be even more important than our calculations show due to the denser oscillation spectrum.

\section{Conclusion}
Our calculations indicate that the many forcing harmonics present in orbits with 
significant eccentricity give rise to effective tidal evolution. By taking into account 
not only the orbital evolution, but also the evolution of the MS star -- especially of 
its rotation rate -- the system progresses through many resonances with stellar 
oscillation modes, a process that drives itself. Frequently conditions are favourable 
for resonance locking, whereby a particular tidal forcing harmonic is kept
near-resonant with an oscillation mode for a prolonged period. Such phases lead to 
rapid orbital and spin evolution. It appears that fairly wide massive binary systems 
with large eccentricity can be almost circularized in a few million years. The observed 
short orbital decay time of the binary pulsar \object{PSR J0045-7319} could be due to 
resonance locking if the stellar spin is retrograde. 

\begin{acknowledgements}
  This work was sponsored by the Stichting Nationale
  Computerfaciliteiten (National Computing Facilities Foundation, NCF)
  for the use of supercomputing facilities, with financial support
  from the Netherlands Organisation for Scientific Research (NWO) and by NWO
  Spinoza grant 08-0 to E.P.J.~van~den~Heuvel.
\end{acknowledgements}


\clearpage
\appendix
\section{Determination of the tidal torque}
\begin{figure}
  \includegraphics{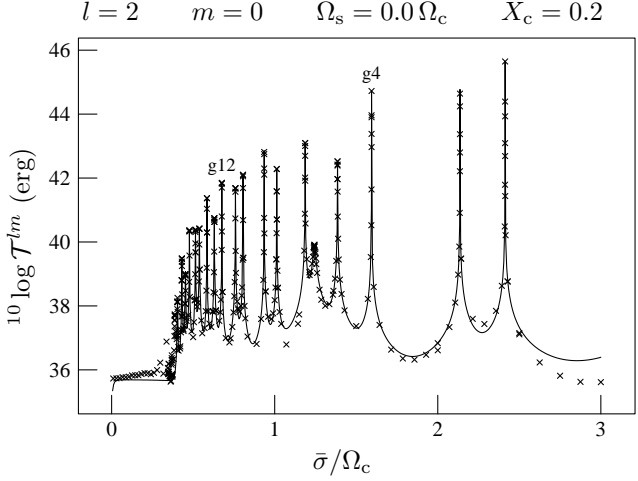}
  \caption[]{Torque integral $\mathcal{T}^{lm}$
    versus forcing frequency $\bar{\sigma}$ for forcing with $l=2$ and $m=0$ on a
    10~$\mathrm{M}_\odot$ non-rotating star.  Prograde g$^2_k$-modes with $k=4$ and $11$ 
    are 
    labelled.  Crosses denote calculated points, while the solid curve
    represents a fit.}
  \label{fig:tqm0omg0}
\end{figure}
In Paper~I we calculated the non-adiabatic tidal response of a uniformly rotating, 
somewhat evolved ($X_c=0.4$), $10\,\mathrm{M}_\odot$ main sequence star to the 
dominant $l=2$ components of its compact $1\,\mathrm{M}_\odot$ binary companion's 
tidal potential for a wide range of prograde and retrograde forcing frequencies and 
determined the corresponding tidal torques on the MS star. The relative orbit of the 
compact star was assumed circular and with a fixed separation of 4 stellar radii (of 
the MS star). In order to include effects of stellar evolution on the tidal 
evolution of the orbit, we have now calculated additional torque spectra for a 
stellar model with core hydrogen abundance (mass fraction) $X_\mathrm{c} = 0.2$, the 
results of which are listed in Tables~\ref{tab:mmodes} and~\ref{tab:rmodes}.

Both stellar models, comprising 1200 (radial) zones, were calculated using the
stellar evolution code developed by Eggleton~\cite*{E72} and more recently revised
by Pols et al.~\cite*{PT95}, with the addition of recent OPAL opacity
Tables~\cite{IR96}.

The model with $X_\mathrm{c}=0.4$ has a stellar radius equal to
$R_\mathrm{s}=3.825 \times 10^{11}$~cm,
an effective temperature $T_\mathrm{eff}=2.314 \times 10^4$~K and a stellar
moment of inertia $I_\mathrm{s}=1.55\times 10^{56} \mbox{ g cm$^2$}$.  The break-up
angular speed equals $\Omega_c=1.54 \times 10^{-4} \mbox{ s$^{-1}$}$.  The model 
with
$X_\mathrm{c}=0.2$, on the other hand, has a radius $R_\mathrm{s}=4.807\times 10^{11}$~cm, an
effective temperature $T_\mathrm{eff} = 2.157\times 10^4$~K, moment of inertia
$I_\mathrm{s}=2.059\times 10^{56}$~g~cm$^2$ and break-up angular frequency
$\Omega_\mathrm{c}=1.095\times 10^{-4}$~s$^{-1}$.

Fig.~\ref{fig:tqm0omg0}
shows the calculated torque integral spectrum for $m=0$ forcing on the non-rotating
star; compared to the $X_\mathrm{c} = 0.4$ data, the degree of central condensation
in the star has increased, the resonance frequencies of the stellar g-modes shift to
much higher dimensionless values $\bar\sigma_0/\Omega_\mathrm{c}$.  The break-up frequency
$\Omega_\mathrm{c}$ however, drops due to the increasing stellar radius, such that
the actual frequency of the modes on average increases by only a few ($\simeq 5$)
percent.  It is also
seen that the spacing between the individual modes is very different for the two
stellar models.  Different modes react differently upon the changes in the stellar
structure, resulting in seemingly random shifting of the peaks superimposed on the
global trend towards higher dimensionless frequencies.

For $X_\mathrm{c}=0.2$ we synthesise g-mode spectra for $\Omega_\mathrm{s} = 0.2$,
$0.3$ and~$0.4\,\Omega_\mathrm{c}$ by normalising the peak widths and areas from the
$m=0$, $\Omega_\mathrm{s}=0.0\,\Omega_\mathrm{c}$, $X_\mathrm{c} = 0.4$ spectrum
from Paper~I to the $m=0$, $\Omega_\mathrm{s}=0.0\,\Omega_\mathrm{c}$ spectrum for
$X_\mathrm{c} = 0.2$ and assuming that the deducted ratios apply to the other
stellar rotation rates as well.  We also assume that the coefficients $c_j$ for the
$\Omega_\mathrm{s}$-dependence $\frac{\bar{\sigma}_0}{\Omega_\mathrm{c}} = \sum c_j
\left(\frac{\Omega_\mathrm{s}}{\Omega_\mathrm{c}}\right)^j$ of the resonance
frequencies calculated for the $X_\mathrm{c} = 0.4$ model in Paper~I can be used for
the $X_\mathrm{c} = 0.2$ spectra.

At low forcing frequencies where the resonant g-mode oscillations attain high radial
orders the values of $\mathcal{T}^{lm}$ are dominated by the 
turbulent dissipation in convective regions (see Paper~I).  For the orbital
evolution calculations of this paper we approximate the torque integral in these
frequency regions by a small constant level. For the non rotating $X_\mathrm{c} = 0.4$
model this level is $1\times10^{37}$~erg, for the $X_\mathrm{c} = 0.2$ model it is
$5\times 10^{35}$~erg. In a rotating model the calculated torque values in the retrograde frequency
domain around the r-modes become larger. Here we multiply the level with a factor
$(1+10\Omega_\mathrm{s}/\Omega_\mathrm{c})$ to approximately fit this effect.
The area $\Delta\bar{\sigma}\mathcal{T}^{lm}$ of this low frequency interval of
the stellar spectrum is not very large, and compares to the peak area of a g-mode
with approximately 15 radial nodes; much less than the peak area of the low radial order
g-modes (1 to 10 radial nodes). Under many circumstances, this regime will
therefore not play an important role in the evolution of the orbit.
Under some circumstances however, e.g.\ if many strong tidal harmonics have frequencies
in the designated area, or if only few harmonics are present in the forcing potential
and the strongest one has its frequency in this domain (as is the case during resonance
locking in moderately eccentric orbits, see Sect.~\ref{sec:lock}), the details of our orbital
calculations are subject to uncertainties due to the numerical limitation of having to
work with a finite mesh and due to the theoretical limitation of not having a sound
theory for turbulent convection. Nevertheless, the calculations presented here are believed
to give a qualitative picture of the process of tidal evolution and are useful as an
order of magnitude estimate for the strength of the occurring effects.

\subsection{Interpolation procedure for the tidal torque}
\label{sec:inphys}

For each forcing frequency $\bar{\sigma}_n$, the corresponding torque integral
$\mathcal{T}_{n}^{lm}$ is calculated by interpolation in the Tables of~\ref{sec:app2}
and those in the appendix of Paper~I in terms of the three state parameters:
forcing frequency $\bar{\sigma}_n$, stellar rotation rate $\Omega_s$ and
evolutionary state of the star $X_\mathrm{c}$ (i.e.\ the hydrogen content of the convective
core).  Subsequently the obtained torque values are normalised by a factor
$(M_\mathrm{p}/\mathrm{M}_\odot)^2(4R_\mathrm{s}/a)^{(2l+2)}(h_n^{(l+1),m})^2$ as
the torque integrals have been calculated for a $1\,\mathrm{M}_\odot$ companion in a
fixed circular orbit with $a=4R_\mathrm{s}$.

To perform interpolations in $\Omega_\mathrm{s}$, linear relations were used for
$\Delta\bar{\sigma}$ and the peak area $\pi \Delta\bar{\sigma} \mathcal{T}^{lm}_0$,
while polynomial interpolations of the highest possible order were used for the
resonance frequencies~$\bar{\sigma}_0$.  For $\bar{\sigma}_0$, we assumed the
derivative $\frac{\mathrm{d}\bar{\sigma}_0}{\mathrm{d}\Omega_\mathrm{s}}$ for $m=0$
modes to vanish for a non-rotating star (from symmetry), and for r-modes to become
$-\frac{1}{3}$ in order to obey the asymptotic value $\bar{\sigma} =
-\frac{2m}{l(l+1)} \Omega_\mathrm{s}$ with $l=3$ and $m=2$.  The r-mode resonance
frequencies were fitted by $\frac{\bar{\sigma}_0}{\Omega_\mathrm{s}} = -\frac{1}{3}
x + c_2 x^2 +c_3 x^3 + c_4 x^4 + c_5 x^5$ with
$x=\Omega_\mathrm{s}/\Omega_\mathrm{c}$, whereby the polynomial coefficients are
listed in Table~\ref{tab:rcf}.

Linear interpolations are used between the $X_\mathrm{c}=0.4$ and~$X_\mathrm{c}=0.2$
models for the peak widths and areas and the frequency $\bar{\sigma}_0$ and
polynomial coefficients $c_j$ of each mode $n$ in the spectra, and for the stellar
radius and moment of inertia.

As the calculated resonance peaks are accurately represented by the harmonic
oscillator fits, the peak width of the fits can be assumed to give a good measure of
the damping timescale $\tau_\mathrm{d} = \frac{2\pi}{\Delta \bar{\sigma}}$ of the
excited modes.  From the tabulated peak widths we find that for all calculated
g-modes $\tau_\mathrm{d}$ lies between $\sim 45$~days and $\sim 10^3$~yr; for the
r-modes $\tau_\mathrm{d}$ varies between $\sim 200$~days and $\sim 200$~years unless
the stellar rotation drops below $0.1\,\Omega_\mathrm{c}$, in which case both the
strength of the modes and the damping rate is strongly reduced.

While evolving the orbit, timesteps are chosen in such a manner that no frequency 
component 
$\bar{\sigma}_n$ will travel more
than what corresponds to five percent of the height of a resonance peak per timestep.

\onecolumn
\setlength{\arraycolsep}{5pt}
\section{Tables}
\label{sec:app2}
\begin{table*}[ht!]
  \caption[]{Fitting parameters for g$^2_k$-mode resonances with an $l=2$, $m=0$
    tidal potential in a $10\,\mathrm{M}_\odot$ star with $X_\mathrm{c}=0.2$. All 
    tabulated values for the
    torque integral have been obtained for a fixed orbital separation.}
  \( 
  \begin{array}{llrl@{\quad}llrl@{\quad}llrl}
    \hline &
    \multicolumn{11}{l}{\Omega_\mathrm{s}=0.0\,\Omega_\mathrm{c}} \\
    k &
    \multicolumn{1}{l}{\bar{\sigma}_0/\Omega_\mathrm{c}} &
    \multicolumn{1}{l}{\mathcal{T}^{lm}_0 \mbox{~(erg)}} &
    \multicolumn{1}{l}{\Delta\bar{\sigma}/\Omega_\mathrm{c}} &
    k &
    \multicolumn{1}{l}{\bar{\sigma}_0/\Omega_\mathrm{c}} &
    \multicolumn{1}{l}{\mathcal{T}^{lm}_0 \mbox{~(erg)}} &
    \multicolumn{1}{l}{\Delta\bar{\sigma}/\Omega_\mathrm{c}} &
    k &
    \multicolumn{1}{l}{\bar{\sigma}_0/\Omega_\mathrm{c}} &
    \multicolumn{1}{l}{\mathcal{T}^{lm}_0 \mbox{~(erg)}} &
    \multicolumn{1}{l}{\Delta\bar{\sigma}/\Omega_\mathrm{c}} \\
    \hline \rule{0mm}{3.5mm}
    1  &  3.4254 &  1.23\times 10^{46} & 1.14\times 10^{-5} & 8  &  1.0127 &  1.94\times 10^{42} & 2.06\times 10^{-4} & 15 &  0.5382 &  2.63\times 10^{40} & 8.91\times 10^{-4} \\
    2  &  2.4130 &  4.53\times 10^{45} & 1.16\times 10^{-5} & 9  &  0.9356 &  7.18\times 10^{42} & 1.00\times 10^{-4} & 16 &  0.5142 &  2.45\times 10^{40} & 7.41\times 10^{-4} \\
    3  &  2.1362 &  5.93\times 10^{44} & 2.93\times 10^{-5} & 10 &  0.8057 &  1.28\times 10^{42} & 1.81\times 10^{-4} & 17 &  0.4768 &  2.30\times 10^{40} & 5.76\times 10^{-4} \\
    4  &  1.5940 &  5.29\times 10^{44} & 2.34\times 10^{-5} & 11 &  0.7592 &  4.86\times 10^{41} & 2.97\times 10^{-4} & 18 &  0.4528 &  1.21\times 10^{39} & 1.91\times 10^{-3} \\
    5  &  1.3853 &  3.32\times 10^{42} & 4.26\times 10^{-4} & 12 &  0.6759 &  7.13\times 10^{41} & 1.64\times 10^{-4} & 19 &  0.4309 &  3.13\times 10^{39} & 8.79\times 10^{-4} \\
    6  &  1.2434 &  7.90\times 10^{39} & 1.48\times 10^{-2} & 13 &  0.6298 &  5.57\times 10^{40} & 6.70\times 10^{-4} & 20 &  0.4039 &  1.72\times 10^{38} & 1.69\times 10^{-3} \\
    7  &  1.1867 &  1.26\times 10^{43} & 2.49\times 10^{-4} & 14 &  0.5846 &  2.35\times 10^{41} & 2.43\times 10^{-4} &    &         &                     &                    \\
    \hline
  \end{array}
  \label{tab:mmodes}
  \)
\end{table*}
\begin{table*}[ht!]
  \caption[]{Fitting parameters for r$^3_k$-mode resonances with an $l=2$, $m=2$
    tidal potential in a 10 $\mathrm{M}_\odot$ star with $X_\mathrm{c}=0.2$. All tabulated values 
    for the torque integral
    have been obtained for a fixed orbital separation.}
  \(
  \begin{array}{l@{\quad}lrl@{\quad}lrl}
    \hline &
    \multicolumn{3}{l}{\Omega_\mathrm{s}=0.1\,\Omega_\mathrm{c}} &
    \multicolumn{3}{l}{\Omega_\mathrm{s}=0.2\,\Omega_\mathrm{c}} \\
    k &
    \multicolumn{1}{l}{\bar{\sigma}_0/\Omega_\mathrm{c}} &
    \multicolumn{1}{l}{\mathcal{T}^{lm}_0 \mbox{~(erg)}} &
    \multicolumn{1}{l}{\Delta\bar{\sigma}/\Omega_\mathrm{c}} &
    \multicolumn{1}{l}{\bar{\sigma}_0/\Omega_\mathrm{c}} &
    \multicolumn{1}{l}{\mathcal{T}^{lm}_0 \mbox{~(erg)}} &
    \multicolumn{1}{l}{\Delta\bar{\sigma}/\Omega_\mathrm{c}} \\
    \hline \rule{0mm}{3.5mm}
    0  & -3.3263\times 10^{-2} & -5.52\times 10^{40} & 8.80\times 10^{-6} & -6.6207\times 10^{-2} & -7.84\times 10^{40} &  1.15\times 10^{-4} \\
    1  & -3.3245\times 10^{-2} & -1.16\times 10^{41} & 1.33\times 10^{-5} & -6.6068\times 10^{-2} & -6.62\times 10^{40} &  1.70\times 10^{-4} \\
    2  & -3.3183\times 10^{-2} & -9.61\times 10^{40} & 2.13\times 10^{-5} & -6.5662\times 10^{-2} & -1.19\times 10^{41} &  1.28\times 10^{-4} \\
    3  & -3.3111\times 10^{-2} & -1.11\times 10^{41} & 1.41\times 10^{-5} & -6.5083\times 10^{-2} & -9.97\times 10^{40} &  1.24\times 10^{-4} \\
    4  & -3.3016\times 10^{-2} & -5.06\times 10^{40} & 2.32\times 10^{-5} & -6.4357\times 10^{-2} & -5.07\times 10^{40} &  1.71\times 10^{-4} \\
    5  & -3.2896\times 10^{-2} & -2.83\times 10^{40} & 2.86\times 10^{-5} & -6.3476\times 10^{-2} & -2.88\times 10^{40} &  1.96\times 10^{-4} \\
    6  & -3.2756\times 10^{-2} & -1.53\times 10^{40} & 3.48\times 10^{-5} & -6.2463\times 10^{-2} & -1.51\times 10^{40} &  2.52\times 10^{-4} \\
    7  & -3.2597\times 10^{-2} & -9.62\times 10^{39} & 4.16\times 10^{-5} & -6.1355\times 10^{-2} & -9.73\times 10^{39} &  2.86\times 10^{-4} \\
    8  & -3.2427\times 10^{-2} & -5.51\times 10^{39} & 5.33\times 10^{-5} & -6.0210\times 10^{-2} & -5.95\times 10^{39} &  3.30\times 10^{-4} \\
    9  & -3.2249\times 10^{-2} & -2.77\times 10^{39} & 5.89\times 10^{-5} & -5.9058\times 10^{-2} & -3.05\times 10^{39} &  3.67\times 10^{-4} \\
    10 & -3.2061\times 10^{-2} & -7.40\times 10^{38} & 7.54\times 10^{-5} & -5.7878\times 10^{-2} & -8.50\times 10^{38} &  4.14\times 10^{-4} \\
    \hline &
    \multicolumn{3}{l}{\Omega_\mathrm{s}=0.3\,\Omega_\mathrm{c}} & 
    \multicolumn{3}{l}{\Omega_\mathrm{s}=0.4\,\Omega_\mathrm{c}} \\
    k &
    \multicolumn{1}{l}{\bar{\sigma}_0/\Omega_\mathrm{c}} &
    \multicolumn{1}{l}{\mathcal{T}^{lm}_0 \mbox{~(erg)}} &
    \multicolumn{1}{l}{\Delta\bar{\sigma}/\Omega_\mathrm{c}} &
    \multicolumn{1}{l}{\bar{\sigma}_0/\Omega_\mathrm{c}} &
    \multicolumn{1}{l}{\mathcal{T}^{lm}_0 \mbox{~(erg)}} &
    \multicolumn{1}{l}{\Delta\bar{\sigma}/\Omega_\mathrm{c}} \\
    \hline \rule{0mm}{3.5mm}
    0  & -9.8673\times 10^{-2} & -7.42\times 10^{40} & 2.23\times 10^{-4} & -1.3043\times 10^{-1} & -7.38\times 10^{40} & 3.69\times 10^{-4} \\
    1  & -9.8199\times 10^{-2} & -9.27\times 10^{40} & 3.88\times 10^{-4} & -1.2937\times 10^{-1} & -1.08\times 10^{41} & 7.91\times 10^{-4} \\
    2  & -9.6772\times 10^{-2} & -1.18\times 10^{41} & 4.08\times 10^{-4} & -1.2598\times 10^{-1} & -1.33\times 10^{41} & 7.57\times 10^{-4} \\
    3  & -9.4960\times 10^{-2} & -9.47\times 10^{40} & 4.17\times 10^{-4} & -1.2199\times 10^{-1} & -8.80\times 10^{40} & 9.93\times 10^{-4} \\
    4  & -9.2676\times 10^{-2} & -5.45\times 10^{40} & 4.50\times 10^{-4} & -1.1721\times 10^{-1} & -5.28\times 10^{40} & 9.91\times 10^{-4} \\
    5  & -9.0077\times 10^{-2} & -2.73\times 10^{40} & 6.36\times 10^{-4} & -1.1189\times 10^{-1} & -2.81\times 10^{40} & 1.14\times 10^{-3} \\
    6  & -8.7159\times 10^{-2} & -1.53\times 10^{40} & 6.86\times 10^{-4} & -1.0630\times 10^{-1} & -1.39\times 10^{40} & 1.47\times 10^{-3} \\
    7  & -8.4114\times 10^{-2} & -9.92\times 10^{39} & 7.52\times 10^{-4} & -1.0076\times 10^{-1} & -1.01\times 10^{40} & 1.36\times 10^{-3} \\
    8  & -8.1102\times 10^{-2} & -6.13\times 10^{39} & 8.90\times 10^{-4} & -9.5551\times 10^{-2} & -6.59\times 10^{39} & 1.42\times 10^{-3} \\
    9  & -7.8246\times 10^{-2} & -3.40\times 10^{39} & 8.35\times 10^{-4} & -9.0814\times 10^{-2} & -3.84\times 10^{39} & 1.44\times 10^{-3} \\
    10 & -7.5465\times 10^{-2} & -1.01\times 10^{39} & 9.48\times 10^{-4} & -8.6386\times 10^{-2} & -1.17\times 10^{39} & 1.69\times 10^{-3} \\
    \hline
  \end{array}
  \label{tab:rmodes}
  \)
\end{table*}
\begin{table*}[ht!]
  \caption{Coefficients $c_j$ for the $\Omega_\mathrm{s}$-dependence
    $\bar{\sigma}_0 = \sum c_j \left(\frac{\Omega_\mathrm{s}}{\Omega_\mathrm{c}}\right)^j$ of the resonance 
    frequency of r-modes in a $10\,\mathrm{M}_\odot$ star with $X_\mathrm{c}=0.4$ and $X_\mathrm{c}=0.2$.
    ($c_1 = -\frac{1}{3}$)}
  \(
  \begin{array}{lr@{}lr@{}lr@{}lr@{}l}
    \hline 
    & \multicolumn{8}{l}{X_\mathrm{c} = 0.4}\\
    k &
    \multicolumn{2}{l}{c_2} &
    \multicolumn{2}{l}{c_3} &
    \multicolumn{2}{l}{c_4} &
    \multicolumn{2}{l}{c_5} \\
    \hline \rule{0mm}{3.5mm}
    0  &  9.&76\times 10^{-3} & -3.&29\times 10^{-2} &  3.&61\times 10^{-1} & -4.&23\times 10^{-1} \\
    1  &  1.&39\times 10^{-2} & -3.&55\times 10^{-2} &  4.&32\times 10^{-1} & -4.&42\times 10^{-1} \\
    2  &  9.&65\times 10^{-3} &  7.&34\times 10^{-2} &  3.&53\times 10^{-1} & -5.&10\times 10^{-1} \\
    3  &  3.&69\times 10^{-3} &  2.&73\times 10^{-1} & -5.&79\times 10^{-2} & -1.&51\times 10^{-1} \\
    4  &  2.&25\times 10^{-4} &  4.&77\times 10^{-1} & -4.&43\times 10^{-1} &  2.&60\times 10^{-1} \\
    5  &  1.&90\times 10^{-3} &  6.&25\times 10^{-1} & -3.&86\times 10^{-1} & -2.&19\times 10^{-1} \\
    6  & -1.&38\times 10^{-3} &  8.&98\times 10^{-1} & -8.&40\times 10^{-1} & -1.&22\times 10^{-1} \\
    7  & -5.&59\times 10^{-3} &  1.&23               & -1.&59               &  3.&45\times 10^{-1} \\
    8  & -1.&27\times 10^{-2} &  1.&64               & -2.&73               &  1.&33               \\
    9  & -1.&60\times 10^{-2} &  2.&00               & -3.&68               &  1.&99               \\
    10 & -3.&45\times 10^{-2} &  2.&68               & -6.&37               &  5.&16               \\
    \hline &\multicolumn{8}{l}{X_\mathrm{c} = 0.2}\\
    k &
    \multicolumn{2}{l}{c_2} &
    \multicolumn{2}{l}{c_3} &
    \multicolumn{2}{l}{c_4} &
    \multicolumn{2}{l}{c_5} \\
    \hline \rule{0mm}{3.5mm}
    0  & -2.&46\times 10^{-5} & 8.&78\times 10^{-2} & -1.&96\times 10^{-1} &  2.&26\times 10^{-1} \\
    1  &  7.&17\times 10^{-4} & 9.&32\times 10^{-2} & -1.&37\times 10^{-1} &  1.&37\times 10^{-1} \\
    2  &  7.&10\times 10^{-3} & 6.&44\times 10^{-2} &  1.&75\times 10^{-1} & -2.&33\times 10^{-1} \\
    3  &  4.&54\times 10^{-3} & 1.&77\times 10^{-1} &  1.&06\times 10^{-2} & -9.&57\times 10^{-2} \\
    4  &  5.&30\times 10^{-3} & 2.&61\times 10^{-1} &  6.&76\times 10^{-2} & -3.&09\times 10^{-1} \\
    5  &  3.&38\times 10^{-3} & 4.&22\times 10^{-1} & -1.&60\times 10^{-1} & -1.&95\times 10^{-1} \\
    6  &  2.&13\times 10^{-3} & 5.&92\times 10^{-1} & -3.&37\times 10^{-1} & -2.&54\times 10^{-1} \\
    7  &  2.&82\times 10^{-4} & 8.&00\times 10^{-1} & -6.&46\times 10^{-1} & -2.&08\times 10^{-1} \\
    8  & -2.&22\times 10^{-3} & 1.&04               & -1.&09               & -3.&14\times 10^{-2} \\
    9  & -6.&60\times 10^{-3} & 1.&33               & -1.&81               &  4.&81\times 10^{-1} \\
    10 & -1.&22\times 10^{-2} & 1.&66               & -2.&73               &  1.&25               \\
    \hline
  \end{array}
  \label{tab:rcf}
  \)
\end{table*}

\end{document}